\documentclass[journal,10pt]{IEEEtran}

\usepackage{amssymb,amsmath}
\usepackage{cite}
\usepackage{url}
\usepackage{xcolor}
\usepackage{graphicx}
\usepackage{subfigure}
\usepackage{fancyhdr}
\usepackage{mdwmath}
\usepackage{mdwtab}
\usepackage{caption}
\usepackage{amsthm}
\usepackage{bm}
\usepackage{bbm}

\usepackage{booktabs}

\usepackage{algpseudocode}

\usepackage{float}

\usepackage[ruled,vlined]{algorithm2e}
\floatname{algorithm}{Algorithm}

\begin{document}

\title{Vehicle Localization via Cooperative Channel Mapping}

\author{Xinghe~Chu,
        Zhaoming~Lu,
        David~Gesbert,~\IEEEmembership{Fellow,~IEEE},
        Luhan~Wang,
        Xiangming~Wen
\thanks{
Part of this work was accepted for publication in a preliminary
conference paper \cite{9145339}.
}
\thanks{Xinghe Chu, Zhaoming Lu, Luhan Wang and Xiangming Wen are with the Beijing Key Laboratory of Network System Architecture and Convergence, Beijing Laboratory of Advanced Information Networks, Beijing University of Posts and Telecommunications, Beijing, China, e-mails: \{chuxinghe, lzy0372, wluhan, xiangmw\}@bupt.edu.cn. (\em{Corresponding author: Zhaoming Lu}.)}
\thanks{David~Gesbert is with the Communications Systems Department, EURECOM, Sophia Antipolis, France, email: David.Gesbert@eurecom.fr.}
}

\markboth{Vehicle Localization via Cooperative Channel Mapping}%
{Shell \MakeLowercase{\textit{et al.}}: Bare Demo of IEEEtran.cls for IEEE Journals}

\maketitle

\begin{abstract}

This paper addresses vehicle positioning, a topic whose importance has risen dramatically in the context of
future autonomous driving systems. While classical methods
that use GPS and/or beacon signals from network
infrastructure for triangulation tend to be sensitive to
 multi-paths
and signal obstruction, our method exhibits robustness with respect to
 such phenomena. Our approach builds on the recently proposed
Channel-SLAM method which first enabled
leveraging of multi-path so as to improve (single) vehicle positioning.
Here,  we propose a cooperative mapping approach which builds upon the
Channel-SLAM concept, referred to here as Team Channel-SLAM.
 Team Channel-SLAM not only
exploits the stationary nature of many reflecting objects around
the vehicle, but also capitalizes on the multi-vehicle nature of
road traffic. The key intuition behind our method is the exploitation for the first time of the correlation between reflectors
around multiple neighboring vehicles. An algorithm is derived
for reflector selection and estimation, combined with a team
particle filter (TPF) so as to achieve high precision simultaneous
multiple vehicle positioning. We obtain
large improvement over the single-vehicle positioning
scenario, with gains being already noticeable for moderate vehicle densities, such as over $\bm{40\%}$ improvement for a vehicle density as low as 4 vehicles in 132 meters' length road.

\end{abstract}


\IEEEpeerreviewmaketitle

\section{Introduction}

\IEEEPARstart{A}{utonomous} driving is widely considered as a potential way to improve traffic safety and mitigate traffic congestion issues \cite{DT}. Vehicle localization is known to play a crucial role in autonomous driving because stable and accurate positioning is needed to assist perception, trajectory planning, and controlling \cite{kuutti2018survey}. What's more, rapid growth in intelligent traffic system applications also fuels the demand for precision localization and tracking  \cite{ulmschneider2016multipath}.
A variety of mechanisms are typically resorted to in the context of vehicle positioning \cite{wymeersch20175g,choudhary2017distributed,lajoie2020door},
the most common tools include Global Positioning System (GPS), Inertial Measurement Unit (IMU), Radio Detection And Ranging (RADAR) and Light Detection And Ranging (LiDAR).
GPS provides absolute positioning, but its biggest problem is when non-line-of-sight (NLOS) and multi-path conditions arise, as its signal is obstructed by buildings or other objects \cite{groves2013height}, which can result in ten-meter scale positioning errors in GPS-challenged areas, like urban canyons \cite{wymeersch20185g}.
IMU provides motion information, but its positioning error tends to accumulate over time, so that the estimated position may drift away \cite{kuutti2018survey}.
RADAR is usually used for relative positioning, yet it cannot estimate the distance if there are obstacles.
LiDAR can improve the precision of localization, but it is an expensive choice and high-power devices are required. The performance of LiDAR is also sensitive to severe weather conditions like rain and snow \cite{kuutti2018survey}.

In terms of radio-based localization, the development of 5G new radio has brought new opportunities for vehicular positioning and tracking \cite{win2018foundations1,win2018foundations2}.
Radio-based positioning is typically based on the estimation of distance and angle parameters, including angle of arrival (AoA), angle of departure (AoD) and time of arrival (ToA) \cite{Fleury1999Channel,richter2005estimation}.
Large bandwidth and multiple antennas in 5G result in higher angle and time resolution, which is essential for high precision localization \cite{wymeersch20185g}.
The basic time unit in Long Term Evolution (LTE) is ${T_{\rm{s}}} = {1 \mathord{\left/
 {\vphantom {1 {\left( {\Delta {f_{{\rm{ref}}}} \cdot {N_{{\rm{f,ref}}}}} \right)}}} \right.
 \kern-\nulldelimiterspace} {\left( {\Delta {f_{{\rm{ref}}}} \cdot {N_{{\rm{f,ref}}}}} \right)}} \approx 32.55{\rm{ns}}$, where $\Delta {f_{{\rm{ref}}}} = 15 \cdot {10^3}{\rm{Hz}}$ and ${N_{{\rm{f,ref}}}} = 2048$. However the basic time unit in 5G NR is ${T_{\rm{c}}} = {1 \mathord{\left/
 {\vphantom {1 {\left( {\Delta {f_{\max }} \cdot {N_{\rm{f}}}} \right)}}} \right.
 \kern-\nulldelimiterspace} {\left( {\Delta {f_{\max }} \cdot {N_{\rm{f}}}} \right)}} \approx 0.509{\rm{ns}}$, where $\Delta {f_{\max }} = 480 \cdot {10^3}{\rm{Hz}}$ and ${N_{\rm{f}}} = 4096$ \cite{3gpp_TS_36_211,3gpp_TS_38_211}. Hence, time resolution has improved greatly in 5G NR.
Additionally, the introduction of massive Multiple Input Multiple Output (MIMO) arrays can significantly improve angle estimation. The overall high temporal and spatial resolution of millimeter wave (mmWave) MIMO makes it possible to estimate those above parameters jointly with satisfactory accuracy \cite{shahmansoori20155g,shahmansoori2018position}.

When it comes to the positioning problem itself, the wireless research literature already offers a rich body of very interesting contributions \cite{del2017survey,bulusu2000gps,win2018theoretical,Marano2010NLOS,mingyang2008distributed,dammann2015prospects,koivisto2017joint,koivisto2017high,savic2015fingerprinting}.
Unfortunately, classical radio-based localization method is hindered by multi-path propagation that introduces parameter ambiguities and noise. \cite{Marano2010NLOS} presents an Least-Squares Support-Vector Machines (LS-SVM) based NLOS identification and mitigation approach to estimate the position of objects by NLOS path identification and mitigation.
Though the approach is promising, it does not make full use of the multi-path structure.

Other interesting works have considered explicitly the presence of the multi-paths \cite{leitinger2016belief,leitinger2015evaluation,gentner2016multipath,leitinger2015simultaneous}.
SLAM-related methods such as \cite{gentner2016multipath,leitinger2015simultaneous,leitinger2019belief} are particularly promising
for how they are able to leverage  AoA and ToA measurements in a clever combined way.
As a special case, the so-called Channel-SLAM method was recently proposed for vehicular positioning \cite{ulmschneider2016multipath,gentner2016multipath}. The key concept behind Channel-SLAM is that multi-paths are recast as virtual direct line of sight radio links that originate from {\it virtual} additional transmitters, whose position is piece-wise stationary and depends on the shape and position of surrounding urban reflectors (see Fig. 1 for a simplified example).
ToA and AoA estimated from vehicles can then be used to localize the virtual transmitters (VTs) and the vehicle {\em simultaneously}.
A recursive Bayesian filter \cite{sorenson1971recursive} or Rao-Blackwellized particle filter (RBPF) \cite{Casella1996Rao} can be used to solve this joint estimation. However, while the approach is promising, it also exhibits limitations that motivate the research elaborated in this paper.
In particular, the above approach fails to exploit the inherent multi-user nature of vehicular traffic and treats the localization of each vehicle completely independently. This is a gross over-simplification, even in moderately dense
traffic, which actually leads to overlooking an opportunity to much improve the localization accuracy.

In this paper, a multi-vehicle collaborative mapping and positioning
approach is proposed. It is referred to as Team Channel-SLAM
(TCS). The main idea behind TCS is to exploit the multi-vehicle
nature of typical road traffic and the ability for different
vehicles to cooperate with each other so as to improve the
localization performance.
The key intuition behind this contribution is
the recognition that multiple vehicles closely following each
other will exhibit a strongly correlated multi-path structure
because they are likely to share one or two common key
reflectors in the form of nearby buildings. Our key point is that the existence of several common virtual transmitters provides a strong additional structure to the vehicle localization problem, hence leading to improved estimation performance in the context of SLAM-based algorithms, with gains expected to grow with traffic density.
Note that preliminary work capturing some of the ideas was presented to a conference version in \cite{9145339}. This present work, however, extends this earlier work in several ways. Importantly, in the present paper we provide a method to associate common virtual transmitters (CVT) to different neighboring vehicles based on a clustering approach. This method allows to track and update the state of common features dynamically. Secondly, a stochastic batch iteration method based on particle filter called team particle filter (TPF) is introduced in this paper, which allows to improve performance substantially by updating the stochastic batch of CVT particles and vehicle particles iteratively.
 The main contributions of this paper are summarized as follows:

\begin{itemize}
  \item  We propose a joint common virtual transmitter (CVT) and vehicle positioning framework which operates on multiple vehicles simultaneously. A key intermediate tool lies in the formation and maintenance of suitable dynamic clusters of CVTs across multiple vehicles, which is presented in this paper.

  \item We propose a new algorithm coined Team Particle Filter (TPF) to achieve multiple vehicle localization and CVT mapping simultaneously. The algorithm iteratively estimates the location of multiple vehicles and CVTs through vehicle particle filters and CVT particle filters cooperatively.

  \item We verify the proposed localization and mapping system through simulations based on ToA and AoA measurements from the literature. Our numerical results indicate that
  the algorithm
   leads to over $40\% $ improvement over the single-vehicle channel-SLAM situation for a traffic denstiy as low as 4 vehicles in a 132 meter long road. For higher densities, higher gains ensue.

\end{itemize}

The rest of the paper is organized as follows. Section II reviews the related works of the proposed localization system. The system model is described in section III. In section IV, we present a CVT formation and cluster maintenance mechanism, and in section V, TPF is proposed for multiple vehicle localization and CVT Mapping. Simulation results are shown in section VI.

\section{Related Works And Background}
Since the proposed localization system achieves simultaneous multiple vehicle localization and CVT mapping,
this section highlights the existing works on multi-path assisted localization.

Multi-path propagation has long been regarded as a drawback for radio-based localization technologies. A large number of contributions draw their focus on multi-path elimination and LOS path extraction \cite{horiba2015accurate,mingyang2008distributed,jiao2009lcrt}.
However,
from a mere information theory standpoint, multi-path components can bring spatial information for vehicular positioning, which makes it possible to turn them from foe to friend in the context of high-accuracy localization \cite{witrisal2016high}.

\cite{wang2011omnidirectional} presents an omnidirectional NLOS identification and localization scheme.
The authors make use of multi-paths to localize the target nodes by NLOS identification and shared reflecting point detection. \cite{wang2011omnidirectional} utilizes multi-paths for localization and finds the common features among multi-paths, but shared reflecting points do not usually exist among multi-paths, which will limit the performance of localization. However, in Team Channel-SLAM, as will be introduced in section IV-A, the multi-paths are recast into LOS links from virtual transmitters to the  receiver and it is in fact likely that some of the multi-paths will share the same reflectors.
\cite{han2018hidden} utilizes multi-paths to detect the position, driving direction and size of a vehicle whose LOS link is obstructed. The approach in \cite{han2018hidden} is promising that the shape of an unknown vehicle can also be estimated, but
the limitation is that the estimation
is restricted to single time-slot worth of data. However,
 many features (like reflecting planes and virtual transmitters) remain static over multiple time slots, which will further enhance positioning in the presence of noise.

To make efficient use of the location information carried by multi-paths, Channel-SLAM was initially proposed in \cite{gentner2016multipath}. Channel-SLAM estimates the position of the virtual transmitters and the user's position jointly with a simultaneous localization and mapping method.
There are also some very interesting single-vehicle Channel-SLAM contributions.
 \cite{wymeersch20185g} exploits the virtual anchors corresponding to vehicles to achieve downlink vehicular positioning, however, a prior map is needed in this contribution. Without foreknowledge of environment, \cite{mendrzik2018joint} presents a message passing-based estimator which jointly estimates the position and orientation of a mobile terminal, and also the location of reflectors or scatters when there does not exist line-of-sight (LOS) path. However, it needs a joint estimation of AoA, AoD and ToA.
 \cite{palacios2017jade} and \cite{palacios2018communication} also develop the Channel-SLAM method for single-user situation without any preknowledge of the environment, but they both need the existence and identification of the LOS path.
The mmWave imaging technology is used for reflected path in \cite{aladsani2019leveraging} to achieve Channel-SLAM, but \cite{aladsani2019leveraging} does not consider the dynamic nature of reflectors for moving objects. \cite{yassin2018mosaic} utilize a non-Bayesian estimator and an extended Kalman filter for single-user Channel-SLAM that can be adopted to an unknown environment, but \cite{yassin2018mosaic} does not consider the association of observations between different time slots.
\cite{kim20205g} proposes a cooperative positioning and mapping approach based on multi-model probability hypothesis density (PHD) filter and map fusion, which is an interesting method for multiple vehicle localization. However, \cite{kim20205g} needs the estimation of AoD and the knowledge of a {\it measurement probability} additionally, where the later indicates how likely it is for a VT with a certain location to give rise to measurements (referred to as ToA, AoA and AoD) to a vehicle in a certain position.
Unfortunately, the constraint on the {\it measurement probability} knowledge limits the environmental adaptability of the algorithm because that distribution will change over time due to the moving of vehicles and changing of the reflectors.
Note that the Team Channel-SLAM goes further than all the above works by exploiting the common virtual transmitters among multiple moving vehicles with changing reflectors.
This is thanks to the fact that multiple neighboring vehicles will share one or more reflectors so that the VTs can be observed through several parallel implicit models, and the multiple observations of the same VT will lead to a more accurate position estimation.

\section{SYSTEM MODEL}
\subsection{Vehicle and Virtual Transmitter Modeling}
We consider a scenario with one base station (BS) and $M$ vehicles. Each vehicle is equipped with one radio user equipment (UE). The base station acts as the only transmitter that transmits radio signals to vehicles, and the vehicles act as $M$ receivers that receive signals from the base station. The state of vehicle at time ${t_k}$ is denoted as
\begin{equation}
\bm{{\cal X}}\left( {{t_k}} \right) = \left\{ {{\bm{x}_1}\left( {{t_k}} \right),...{\bm{x}_m}\left( {{t_k}} \right),...,{\bm{x}_M}\left( {{t_k}} \right)} \right\}\label{state_veh}
\end{equation}
\noindent where ${\bm{x}_m}\left( {{t_k}} \right)$ is the state for $m$-th vehicle, $m = 1,...,M$,
\begin{equation}
  {\bm{x}_m}\left( {{t_k}} \right) = \left\{ {\bm{r}_{{V_m}}\left( {{t_k}} \right),\bm{v}_m\left( {{t_k}} \right)} \right\}\label{x_m_define}
\end{equation}
where ${\bm{r}_{{V_m}}}\left( {{t_k}} \right) \in {{\rm{R}}^2}$ and $\bm{v}_m\left( {{t_k}} \right) \in {{\rm{R}}^2}$ are its position and velocity, respectively.

\begin{figure}[t]
\centerline{\includegraphics[width=6cm]{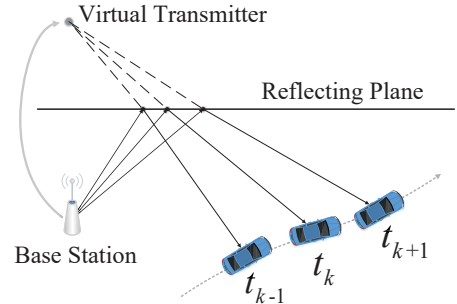}}
\caption{The signal from the base station is reflected and then received by a vehicle-borne UE.
Then a virtual transmitter can be seen locating at the mirror position of the transmitter (BS) over the reflecting plane sending signals to the moving UE through a LOS link.
}
\label{VT_model}
\end{figure}

As shown in Fig. \ref{VT_model}, when there is a reflecting path between the base station and the $m$-th vehicle at time slot $t_k$, the multi-path can be recast into a LOS link between the $m$-th vehicle and a VT, who locates in the mirror position of the base station over the reflecting plane and is static over some time \cite{gentner2016multipath}. The state of VTs observed by each vehicle at time slot $t_k$ is denoted as:
\begin{equation}
\bm{{\cal V}}\left( {{t_k}} \right) = \left\{ {{\bm{r}_{VT_1}}({t_k}),...,{\bm{r}_{VT_m}}({t_k}),...,{\bm{r}_{VT_M}}\left( {{t_k}} \right)} \right\}\label{state_vt}
\end{equation}
where ${{\bm{r}_{VT_m}}({t_k})},m = 1,2,...M$ denote the position of the VTs for $m$-th vehicle,
\begin{equation}
  {\bm{r}_{VT_m}}({t_k}) = \left\{ {{\bm{r}_{VT_{\left( {m,{p_m}} \right)}}}({t_k})} \right\}_{{p_m} = 1}^{{N_m}\left( {{t_k}} \right)}\label{vt_def}
\end{equation}
${N_m}\left( {{t_k}} \right)$ denote the number of multi-paths observed by the $m$-th vehicle at time slot ${t_k}$, and ${\bm{r}_{VT\left( {m,{p_m}} \right)}}\left( {{t_k}} \right) \in {{\rm{R}}^{\rm{3}}}$ is the position of the VT recast by the $p_m$-th multi-path from the $m$-th vehicle at time slot ${t_k}$, which can be calculated by (\ref{cal_VTs}) in subsection B.

Note that for vehicle localization, since the height of vehicles is constant over time, only a 2D scenario is considered. However, for VT (or CVT later in the paper) localization, 3-D positioning is considered.

\begin{figure*}[ht]
\centering
\subfigure[The Common Physical Transmitter Model. ]{\includegraphics[width=9cm]{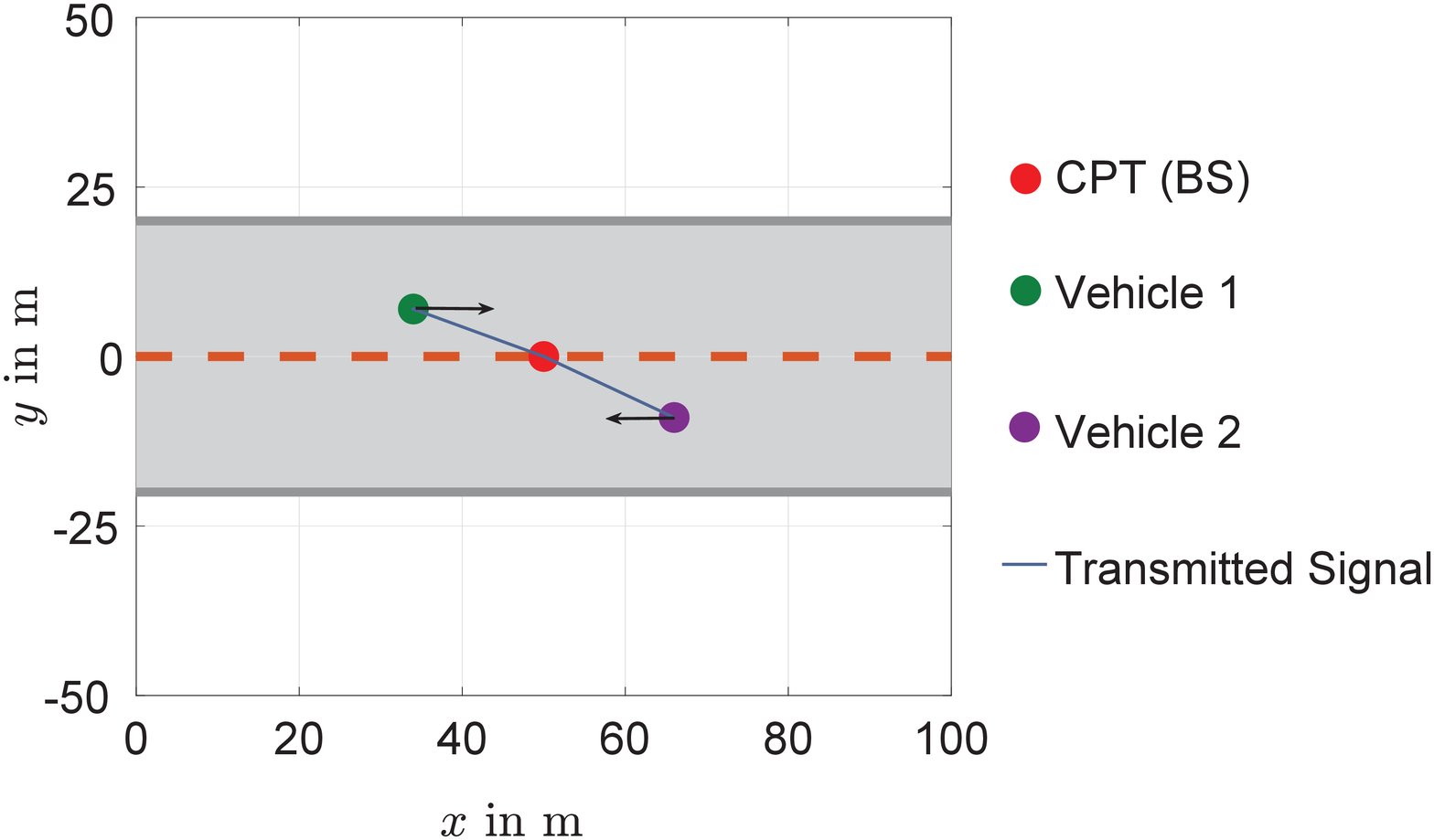}}%
\subfigure[The Common-Reflector Virtual Transmitter Model.]{\includegraphics[width=9cm]{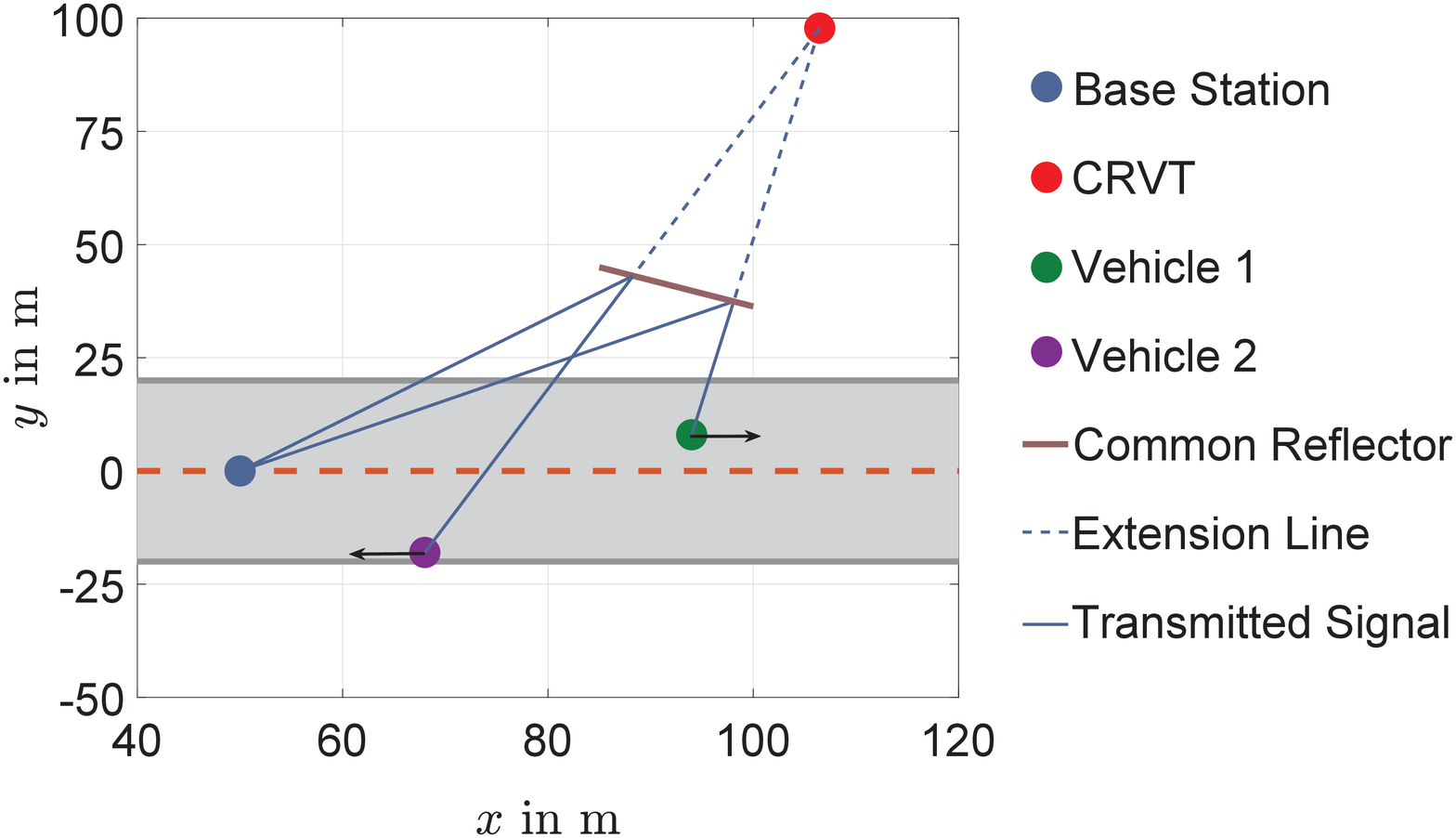}}%
\caption{The Common Virtual Transmitter model includes the common physical transmitter model and the common-reflector virtual transmitter model.}
\label{CVT_model}
\end{figure*}

\subsection{Multi-Path Observation}
The measurements from multi-paths include ToA and AoA. They are assumed to be given, and the particular choice of method used to obtained such measurements is left unspecified \cite{blanco2019performance}.
The observation at time slot ${t_k}$ is denoted as
\begin{equation}
  \begin{array}{c}
{\cal Z}\left( {{t_k}} \right) = \left\{ {{z_1}\left( {{t_k}} \right),...,{z_m}\left( {{t_k}} \right),...,{z_M}\left( {{t_k}} \right)} \right\} \vspace{0.5ex} \\
  {z_m}\left( {{t_k}} \right) = \left\{ {{z_{\left( {m,{p_m}} \right)}}\left( {{t_k}} \right)} \right\}_{{p_m} = 1}^{{N_m}\left( {{t_k}} \right)}
\end{array} \label{z_m_def}
\end{equation}
where ${{z_m}\left( {{t_k}} \right)}$ denotes all the multi-path measurements from $m$-th vehicle, and ${{z_{\left( {m,{p_m}} \right)}}\left( {{t_k}} \right)}$ is the $p$-th path in ${{z_m}\left( {{t_k}} \right)}$ denoting as
\begin{equation}
  {z_{\left( {m,{p_m}} \right)}}\left( {{t_k}} \right) = \left\{ {{\bm{{\widehat \alpha }}_{\left( {m,{p_m}} \right)}}\left( {{t_k}} \right),{{\widehat d}_{\left( {m,{p_m}} \right)}}\left( {{t_k}} \right)} \right\}\label{z_m_im_def}
\end{equation}
${\bm{\widehat \alpha} _{\left( {m,{p_m}} \right)}}\left( {{t_k}} \right) = \left( {{{\widehat \theta }_{\left( {m,{p_m}} \right)}},{{\widehat \varphi }_{\left( {m,{p_m}} \right)}}} \right)$ is the estimation of AoA, where  ${{{\widehat \theta }_{\left( {m,{p_m}} \right)}}}$ and ${{{\widehat \varphi }_{\left( {m,{p_m}} \right)}}}$ denote the polar angle and azimuth angle of AoA, respectively.
${{{\widehat d}_{\left( {m,{p_m}} \right)}}\left( {{t_k}} \right)}$ is the ToA estimation multiplied by the light speed.

So the position of the VT recast from the multi-path measured as ${{z_{\left( {m,{p_m}} \right)}}\left( {{t_k}} \right)}$ is calculated as
\begin{equation}
  {\bm{r}_{VT\left( {m,{p_m}} \right)}}{\rm{ = }}{\bm{r}_{{V_m}}}\left( {{t_k}} \right) + \overrightarrow R \left( {{z_{\left( {m,{p_m}} \right)}}\left( {{t_k}} \right)} \right) \label{cal_VTs}
\end{equation}
where $\overrightarrow R \left( {\bm{\alpha} ,d} \right)$ is the  vector operator, which is defined as $\overrightarrow R \left( {\bm{\alpha} ,d} \right) = d \cdot \left( {\sin \left( \varphi  \right)\cos \left( \theta  \right),\sin \left( \varphi  \right)\sin \left( \theta  \right),cos\left( \varphi  \right)} \right)$.

\subsection{Problem Formulation}
A SLAM based method is introduced to solve the vehicle localization problem.
SLAM is a map-based localization and tracking method without prebuilt map \cite{bresson2017simultaneous,durrant2006simultaneous}. Specifically, in feature-based SLAM \cite{dissanayake2001solution,mullane2011random}, the map is made up of an unknown number of features, whose state is estimated simultaneously with the objects based on sequential measurements from sensors. Consequently, SLAM estimates the state of targets and the map features around those targets by sensors like camera, LiDAR or radio signal, where the localization process and the mapping process are ongoing simultaneously. Mathematically, SLAM computes the following probability distribution
\begin{equation}
{\rm{p}}\left( {\bm{x}\left( {{t_k}} \right),\bm{m}\left| {z\left( {{t_{1:k}}} \right)} \right.,u\left( {{t_{1:k}}} \right),\bm{x}\left( {{t_0}} \right)} \right) \label{SLAM}
\end{equation}
where $\bm{x}$ refers to the state (e.g. position) of the target, $\bm{m}$ contains the features of the map, $z$ stands for observations from sensors, and $u$ stands for the motion information between different time slots.

The Channel-SLAM and Team Channel-SLAM borrow from the SLAM principles to solve a new problem which is that of vehicle positioning, where the localization refers to vehicle localization, and the mapping refers to the positioning of VTs (subliming to CVT in Team Channel-SLAM). The localization and mapping process are executed simultaneously in this paper for high accuracy vehicle localization. In detail, the CVT positioning process is achieved based on the observation of multi-path as well as the vehicle localization process, and the CVT positioning process can also assist with the vehicle localization process in return to improve the positioning accuracy.

In the Channel SLAM context, the probability distribution turns into
\begin{equation}
{\rm{p}}\left( {\bm{x}\left( {{t_k}} \right),\bm{{\cal V}}\left| {z\left( {{t_{1:k}}} \right)} \right.,u\left( {{t_{1:k}}} \right),\bm{x}\left( {{t_0}} \right)} \right) \label{channel-SLAM}
\end{equation}
where $\bm{{\cal V}}$ stands for the state of virtual transmitters.

Since in Team Channel-SLAM, localization simultaneously targets multiple vehicles, the probability distribution evolves into
\begin{equation}
{\rm{p}}\left( {\underbrace {\bm{{\cal X}}\left( {{t_k}} \right)}_{{\bm{x}_1},...,{\bm{x}_M}},\bm{{\cal C}}\left| {\underbrace {{\cal Z}\left( {{t_{1:k}}} \right)}_{{z_1},...,{z_M}}} \right.,\underbrace {{\cal U}\left( {{t_{1:k}}} \right)}_{{u_1},...,{u_M}},\underbrace {\bm{{\cal X}}\left( {{t_0}} \right)}_{{\bm{x}_1},...,{\bm{x}_M}}} \right)\label{Team Channel-SLAM}
\end{equation}
where $M$ is the number of vehicles, and $\bm{{\cal C}}$ stands for state of common virtual transmitters, which will be introduced in next section.

\section{CVT Formation and Cluster Maintenance}

\subsection{Common Virtual Transmitter model}

When there are $M$ vehicles in the network, there would be totally $\sum\limits_{m = 1}^M {{N_m}\left( {{t_k}} \right)} $ paths at time slot $t_k$. On the one hand, the base station is the common origin to all the paths. On the other hand, some multi-paths may be reflected by the same reflector, thereby some groups of VTs may have the same position in theory. Such a group of VTs will be lumped into what we call a Common Virtual Transmitter (CVT).
As shown in Fig.\ref{CVT_model}, there are two kinds of CVT.

\textbf{a) Common Physical Transmitter:} This refers to the case where the paths originate directly from the same physical transmitter (here the BS) in LOS fashion as shown in Fig.\ref{CVT_model} (a).

\textbf{b) Common-Reflector Virtual Transmitter:} As shown in Fig.\ref{CVT_model} (b), when multi-paths share the same reflector, the VTs observed by these multi-paths can be seen as a CVT with a unique position.

\subsection{Affinity Propagation Based Clustering}
The VTs treated as one CVT are usually not
completely identical because the noise corrupts the observations.
However, they are expected to be close to each other.
This notion of correlated VT can be articulated via a concept of VT {\it cluster}, which is now described.
The key advantage of clustering together the common (or near common) virtual transmitters that associate with multiple vehicles is that a better robustness to noise is obtained in the estimation process.

Affinity propagation \cite{frey2007clustering,liu2017distributed} is a popular clustering method that can collect a group of nodes by choosing an exemplar (cluster head) for each node without preknowledge of the number of clusters and their size.
For clustering purposes, we need to introduce the negative logarithm of distance between VTs defined below, which is referred to later on as {\it similarity} value.
\begin{equation}
  \begin{array}{*{20}{c}}
  {s\left( {\bm{p},\bm{q}} \right) =  - \ln \left( {\left\| {{\bm{r}_{V{T_{\bm{p}}}}} - {\bm{r}_{V{T_{\bm{q}}}}}} \right\| + 1} \right)} \vspace{0.5ex} \\
  {\bm{p} \ne {\bm{q}},\;{\bm{r}_{V{T_{\bm{p}}}}},{\bm{r}_{V{T_{\bm{q}}}}} \in \left\{ {\left\{ {{\bm{r}_{VT\left( {m,{p_m}} \right)}}} \right\}_{{p_m} = 1}^{{N_m}\left( {{t_k}} \right)}} \right\}_{m = 1}^M}
  \end{array}\label{model_similarity}
\end{equation}
where $s\left( {\bm{p},{\bm{q}}} \right)$ is the similarity value of VT $\bm{p}$ with respect to VT ${\bm{q}}$, which indicates how well VT ${\bm{q}}$ is suited
to be the exemplar for VT $\bm{p}$ \cite{frey2007clustering}. Note that ${\bm p}$ and ${\bm q}$ are the corresponding multi-path index in equation (\ref{model_similarity}).

The affinity propagation algorithm continuously updates and iterates a so-called {\it responsibility} value $r\left( {\bm{p},{\bm{q}}} \right)$ and {\it availability} value $a\left( {\bm{p},{\bm{q}}} \right)$, where the former means the accumulated evidence for how well-suited VT
${\bm{q}}$ can serve as the exemplar for VT $\bm{p}$ and the latter means the accumulated evidence for how
appropriate it would be for VT $\bm{p}$ to choose
VT ${\bm{q}}$ as its exemplar. How we obtain such values iteratively is shown below in equations (\ref{res_cal}$\sim$\ref{ava_damp}).
 Finally, for any VT $\bm{p}$, the VT ${\bm{q}}$ that maximizes the sum of {\it responsibility} value and {\it availability} value
is selected as an exemplar of VT $\bm{p}$ \cite{frey2007clustering}. One iteration of affinity propagation for CVT cluster formation is done as follows:

\textbf{a) Calculate the responsibility value:}
\begin{equation}
r\left( {\bm{p},{\bm{q}}} \right) \leftarrow s\left( {\bm{p},{\bm{q}}} \right) - \mathop {\max }\limits_{{\bm{q}}'{\rm{s}}.{\rm{t}}.{\bm{q}}' \ne {\bm{q}}} \left\{ {a\left( {\bm{p},{\bm{q}}'} \right) + s\left( {\bm{p},{\bm{q}}'} \right)} \right\} \label{res_cal}
\end{equation}

\textbf{b) Damp the responsibility value:}
\begin{equation}
r\left( {\bm{p},{\bm{q}}} \right) \leftarrow \left( {1 - \lambda } \right)r\left( {\bm{p},{\bm{q}}} \right) + r{\left( {\bm{p},{\bm{q}}} \right)^{old}} \label{res_damp}
\end{equation}
\noindent In equation (\ref{res_damp}), $\lambda $ is a damping factor between 0 and 1, which is set to 0.9 to gives a good compromised between the effect of eliminating oscillations and convergence rate according to \cite{han2019production}. $r{\left( {\bm{p},{\bm{q}}} \right)^{old}}$ is the {\it responsibility} value at previous iteration.

\textbf{c) Calculate the availability value:}
\begin{equation}
a\left( {\bm{p},{\bm{q}}} \right) \leftarrow \min \left\{ {0,r\left( {{\bm{q}},{\bm{q}}} \right) + \sum\limits_{{\bm{p}^\prime } \notin \left\{ {\bm{p},{\bm{q}}} \right\}} {\max \left\{ {0,r\left( {\bm{p}',{\bm{q}}} \right)} \right\}} } \right\}\label{ava_cal}
\end{equation}

\textbf{d) Damp the availability value:}
\begin{equation}
a\left( {\bm{p},{\bm{q}}} \right) \leftarrow \left( {1 - \lambda } \right)a\left( {\bm{p},{\bm{q}}} \right) + a{\left( {\bm{p},{\bm{q}}} \right)^{old}} \label{ava_damp}
\end{equation}
$\lambda$ is the damping factor the same as equation (\ref{res_damp}), and $a{\left( {\bm{p},{\bm{q}}} \right)^{old}}$ is the availability value at previous iteration.

\textbf{e) Calculate the self-availability value:}
\begin{equation}
a\left( {{\bm{q}},{\bm{q}}} \right) \leftarrow \sum\limits_{{\bm{p}^\prime } \ne {\bm{q}}} {\max \left\{ {0,r\left( {\bm{p}',{\bm{q}}} \right)} \right\}} \label{self_ava_cal}
\end{equation}

\textbf{f) Choose exemplar:}
\begin{equation}
\left\{ \begin{array}{l}
{\rm{if}}\;\;iter > {N_{iter}}\;\;{E_{\bm{p}}} = \mathop {\arg \max }\limits_{\bm{q}} \left\{ {a\left( {\bm{p},{\bm{q}}} \right) + r\left( {\bm{p},{\bm{q}}} \right)} \right\}\\
{\rm{if}}\;\;iter \le {N_{iter}}\;\;{\rm{return}}\;\;{\rm{step\ \textbf{a)}}}\;
\end{array} \right.\label{exemplar_choose}
\end{equation}
where VT ${E_{\bm{p}}}$ is the exemplar for VT $\bm{p}$.

After \textbf{a)}$\;\sim\;$\textbf{f)}, ${N_C}\left( {{t_k}} \right)$ exemplars are chosen among all the VTs at time slot $t_k$.
\begin{equation}
{E\left( u \right),u = 1,...,{N_C}\left( {{t_k}} \right)}\label{e_t_k}
\end{equation}
where $E\left( u \right)$ denotes the $u$-th exemplar.
 Then ${N_C}\left( {{t_k}} \right)$ clusters are selected by grouping the VTs with the same exemplar together.
\begin{equation}
{{\cal{S}}_u}\left( {{t_k}} \right) = \left\{ {{\bm p}\left| {{E_{\bm p}} = E\left( u \right)} \right.} \right\},u = 1,...,{N_C}\left( {{t_k}} \right)\label{c_t_k}
\end{equation}
where ${\bm p}$ is the index of the multi-paths  shaped like $\left( {m,{p_m}} \right)$, and ${{\cal{S}}_u}\left( {{t_k}} \right)$ is the set of multi-paths such that their corresponding VTs have the same exemplar ${E\left( u \right)}$.

The VTs with the same exemplar are declared to be grouped into one CVT cluster.
Thus the state of CVT clusters are denoted as:

\begin{equation}
  \begin{array}{c}
  \bm{{\cal C}}\left( {{t_k}} \right) = \left\{ {{\bm{{C}}_u}\left( {{t_k}} \right)} \right\}_{u = 1}^{{N_C}\left( {{t_k}} \right)} \vspace{0.5ex}\\
  {\bm{C}_u}\left( {{t_k}} \right) = \left\{ {{\bm{r}_{{C_u}}}\left( {{t_k}} \right),{\bm{I}_{{C_u}}}\left( {{t_k}} \right)} \right\}
  \end{array} \label{state_cluster}
\end{equation}
Since each exemplar corresponds to a CVT cluster, so ${N_{C}}\left( {{t_k}} \right)$ can also denote the number of CVT clusters at time slot $t_k$, and ${{\bm{r}_{{C_u}}}\left( {{t_k}} \right)}$ is the 3-D position of CVT, which is the mean value of its corresponding VTs' positions.
\begin{equation}
  \begin{array}{*{20}{l}}
  {{\bm{r}_{{C_u}}}\left( {{t_k}} \right) = \mathbb{E}\left( {{\bm{r}_{V{T_{\bm{p}}}}}\left( {{t_k}} \right)} \right),{\bm{p}} \in {{\cal S}_u}\left( {{t_k}} \right)}
  \end{array}\label{rch_cal}
\end{equation}
where $\mathbb{E}\left(  \cdot  \right)$ is the expectation operator, and ${{r_{V{T_{\bm{p}}}}}\left( {{t_k}} \right)}$ is the position of VTs in cluster $u$.

The common virtual transmitter index vector notation ${\bm{I}_u}\left( {{t_k}} \right)$ is then introduced to indicate whether the $u$-th CVT cluster contains a VT observed by vehicle. If the $m$-th vehicle has an observation to that CVT, then the $m$-th element in ${\bm{I}_u}\left( {{t_k}} \right)$ (denoted as ${\pi_{u,m}}\left( {{t_k}} \right)$) equals to the corresponding multi-path index $p_m$, otherwise ${\pi_{u,m}}\left( {{t_k}} \right)$ is 0.
\begin{equation}
  \begin{array}{c}
  {\bm{I}_{C_u}}\left( {{t_k}} \right) = \left[ {{\pi_{u,1}}\left( {{t_k}} \right),...,{\pi_{u,m}}\left( {{t_k}} \right),...,{\pi_{u,M}}\left( {{t_k}} \right)} \right]\vspace{0.5ex} \\
  {\pi_{u,m}} = \left\{ {\begin{array}{*{20}{l}}
  {{{{\hat p}_m}},\;{\rm{if}}\ \ {{{\hat p}_m}}={{\cal S}_u}\left( {{t_k}} \right) \cap \left\{ {\left( {m,{p_m}} \right)} \right\}_{{p_m} = 1}^{{N_m}\left( {{t_k}} \right)} \ne \phi }\\
  {0,\;\;\ {\rm{otherwise}}}
  \end{array}} \right.
\end{array} \label{model_judge_CVT}
\end{equation}
where ${\left\{ {\left( {m,{p_m}} \right)} \right\}_{{p_m} = 1}^{{N_m}\left( {{t_k}} \right)}}$ denotes all the multi-path indexes of $m$-th vehicle.

\begin{algorithm}[t]
\caption{CVT Formation and Cluster Maintenance}
\LinesNumbered
\KwIn{$vehicle\ state$ : $\bm{{\cal X}}\left( {{t_{k-1}}} \right)$\\
\setlength{\parindent}{3em}$observations$ : ${\cal Z}\left( {{t_k}} \right)$\\
\setlength{\parindent}{3em}$motion\ information$ : ${\cal U}\left( {{t_{k}}} \right)$\\
}
\KwOut{$CVT\ state$ : $\bm{{\cal C}}\left( {{t_k}} \right)$\\
}
Calculate $\bm{{\cal X}}\left( {{t_{k}}} \right)$ based on $\bm{{\cal X}}\left( {{t_{k-1}}} \right)$ and ${\cal U}\left( {{t_{k}}} \right)$ \\
Calculate $\bm{{\cal V}}\left( {{t_k}} \right)$ based on $\bm{{\cal X}}\left( {{t_{k}}} \right)$ and ${\cal Z}\left( {{t_k}} \right)$ as (\ref{cal_VTs})\\
Clustering the CVT as (\ref{model_similarity}$\sim$\ref{model_judge_CVT})\\
New VT joining as (\ref{vt_newly_ob}$\sim$\ref{standalone_cvt_cluster})\\
Neighbor CVT cluster merging as (\ref{Q_M_def}$\sim$\ref{merging_upd})\\
Out-dated CVT deleting as (\ref{cvt_delete})\\
\label{A_1}
\end{algorithm}

\subsection{Dynamic Cluster Maintenance and Association Mechanism}

Due to the dynamic nature of the moving vehicles,
the observed VTs will also be dynamically changed.
 Thus on the one hand, there may possibly arise some VT observations to a CVT that does not exist in the previous time slot. On the other hand, it is also possible that there will no longer exist any VT observation to a certain CVT existing in the previous time slot.
So there exists a birth-death process for CVTs. To this end, when the CVT clusters are established by the algorithm mentioned in section IV-B for one particular time slot, they have to be updated from time slot to time slot based on the newly observed VTs.
In the below, we introduce a mechanism to firstly associate every new VT observation to a certain CVT clusters and then update the CVT cluster dynamically. The main idea behind this section is cluster association and cluster maintenance:
\begin{itemize}
\item {\em Cluster Association:} To associate the newly observed VTs to the current CVT clusters.
\item {\em Cluster Maintenance:} To update the CVT clusters dynamically based on the previously observed VTs and newly observed VTs.
\end{itemize}

The mechanism is denoted as dynamic cluster maintenance and association mechanism, which is described as follows:

\textbf{a) New VT joining:} when a multi-path is newly observed, the position of its corresponding VT is then calculated by (\ref{cal_VTs}).
\begin{equation}
\bm{{\cal V}}\left( {{t_{k + 1}}} \right) = \left\{ {\left\{ {{\bm{r}_{VT\left( {m,{p_m}} \right)}}\left( {{t_{k + 1}}} \right)} \right\}_{{p_m} = 1}^{{N_m}\left( {{t_{k + 1}}} \right)}} \right\}_{m = 1}^M \label{vt_newly_ob}
\end{equation}
The data association quality is defined as :
\begin{equation}
  \begin{array}{c}
  {Q_A}\left( {{\pi_{u,m}}\left( {{t_{k + 1}}} \right) = {p_m}} \right) =  - \ln \left( {\Delta _u^{\left( {m,{p_m}} \right)}\left( {{t_k}} \right)  + 1} \right) \vspace{0.5ex}\\
  \Delta _u^{\left( {m,{p_m}} \right)}\left( {{t_k}} \right) = \left\| {{\bm{r}_{{C_u}}}\left( {{t_k}} \right) - {\bm{r}_{VT\left( {m,{p_m}} \right)}}\left( {{t_{k + 1}}} \right)} \right\|
  \end{array}\label{Q_A_def}
\end{equation}
Where ${Q_A}\left( {{\pi_{u,m}}\left( {{t_{k + 1}}} \right) = {p_m}} \right) $ indicates how close the VT observed by multi-path $\left( {m,{p_m}} \right)$ is to the $u$-th CVT, which further indicates how likely that the VT originated from multi-path $\left( {m,{p_m}} \right)$ belongs to CVT cluster $u$.

For multi-path $\left( {m,{p_m}} \right)$, the CVT cluster that maximizes ${Q_A}\left( {{\pi_{u,m}}\left( {{t_{k + 1}}} \right) = {p_m}} \right) $ is denoted as $u_A$:
\begin{equation}
  {u_A} = \mathop {\arg \max }\limits_u \left\{ {{Q_A}\left( {{\pi_{u,m}}\left( {{t_{k + 1}}} \right) = {p_m}} \right)} \right\}\label{def_u_A}
\end{equation}
Then the common virtual transmitter index vector related to the $m$-th vehicle ${\pi _{u,m}}\left( {{t_k}} \right),u = 1,2,...,{N_C}\left( {{t_k}} \right)$ is updated as:
\begin{equation}
  {\pi _{u,m}}\left( {{t_{k + 1}}} \right) = \left\{ \begin{array}{l}
  {p_m},\;\;\;\;{\rm{if}}\ \ {Q_A}\left( {{\pi _{u,m}}\left( {{t_{k + 1}}} \right) = {p_m}} \right) \ge {L_A}\\
  \ \ \ \ \ \ \ \ \ \ \ \ {\&\&}\ \ u = {u_A}\\
  0,\;\;\;\;\;\;{\rm{otherwise}}
  \end{array} \right.
  \label{1_u_m_upd}
\end{equation}
where $L_A$ is a suitable data association threshold.

If the multi-path $\left( {m,{p_m}} \right)$
 is associated with CVT cluster $u$, the position of CVT cluster $u$ is updated as:
\begin{equation}
  \begin{array}{c}
{{\cal S}_u}\left( {{t_{k + 1}}} \right) = \left( {m,{p_m}} \right) \cup {{\cal S}_u}\left( {{t_k}} \right) \vspace{0.5ex}\\
{\bm{r}_{{C_u}}}\left( {{t_{k + 1}}} \right) = \mathbb{E}\left( {{\bm{r}_{V{T_{\bm{p}}}}}\left( {{t_{k + 1}}} \right)} \right),{\bm{p}} \in {{\cal S}_u}\left( {{t_{k + 1}}} \right)
\end{array}\label{r_u_t_1_upd}
\end{equation}

However, there may exist some multi-paths that cannot associate to any CVT cluster because those multi-paths are reflected new reflectors. In this situation, those multi-paths are placed into standalone CVT clusters. Those CVT clusters are established as follows:
\begin{equation}
  \begin{array}{c}
{\bm{r}_{{C_{u'}}}}\left( {{t_{k + 1}}} \right) = {\bm{r}_{VT\left( {m,{p_m}} \right)}}\left( {{t_{k + 1}}} \right) \vspace{0.5ex}\\
{\pi_{u',m'}}\left( {{t_{k + 1}}} \right) = \left\{ \begin{array}{l}
{p_m},{\ \ \ }\ \ {\rm if}\ m' = m \vspace{0.5ex}\\
0,{\ \ \ \ \ \ \ \rm{otherwise}}
\end{array} \right.
\end{array}\label{standalone_cvt_cluster}
\end{equation}
where we use $u'$ to denote the index for the standalone CVT, and ${\pi _{u',m'}}\left( {{t_{k + 1}}} \right),m' = 1,2,...,M$ is the common virtual transmitter index vector for the $u'$-th CVT.

\textbf{b) Neighbor CVT cluster merging :} when two CVTs are close to each other, they can be merged into one CVT cluster. Then we define the cluster merging quality as:
\begin{equation}
  \begin{array}{c}
{Q_M}\left( {{\bm{C}_i}\left( {{t_k}} \right) , {\bm{C}_j}\left( {{t_k}} \right)} \right) =  - \ln \left( {\Delta _i^j\left( {{t_k}} \right) + 1} \right) \vspace{0.5ex}\\
\Delta _i^j\left( {{t_k}} \right) = \left\| {{\bm{r}_{{C_i}}}\left( {{t_k}} \right) - {\bm{r}_{{C_j}}}\left( {{t_k}} \right)} \right\|
\end{array}\label{Q_M_def}
\end{equation}
Where ${Q_M}\left( {{\bm{C}_i}\left( {{t_k}} \right) , {\bm{C}_j}\left( {{t_k}} \right)} \right)$ indicates how likely CVT cluster $i$ is equivalent to CVT cluster $j$.

For CVT cluster $i$ and CVT cluster $j$, cluster $i$ is equivalent to cluster $j$ if they satisfy the following equation:
\begin{equation}
  \begin{array}{c}
{Q_M}\left( {{\bm{C}_i}\left( {{t_k}} \right),{\bm{C}_j}\left( {{t_k}} \right)} \right) \ge {L_M}\;   \vspace{0.5ex}\\
{\bm{I}_{{C_i}}}\left( {{t_k}} \right) \circ {\bm{I}_{{C_j}}}\left( {{t_k}} \right) = O
\end{array} \label{def_cluster_merging}
\end{equation}
where $L_M$ is a suitable CVT cluster merging threshold and $ \circ $ is the Hadamard product  defined as ${\left( {\bm{A} \circ \bm{B}} \right)_{ij}} = {\left( \bm{A} \right)_{ij}}{\left( \bm{B} \right)_{ij}}$, which prevents that there are more than one multi-path from a vehicle participating in the formation of one CVT cluster.

If CVT cluster $i$ is equivalent to CVT cluster $j$, the merged CVT cluster $u'$ is updated as:
\begin{equation}
  \begin{array}{c}
{\bm{r}_{{C_{u'}}}}\left( {{t_k}} \right) = \mathbb{E}\left( {{\bm{r}_{V{T_{\bm{p}}}}}\left( {{t_k}} \right)} \right),{\bm{p}} \in {{{\cal S}}_i}\left( {{t_k}} \right) \cup {{{\cal S}}_j}\left( {{t_k}} \right)\vspace{0.5ex} \\
{\bm{I}_{{C_{u'}}}}\left( {{t_k}} \right) = {\bm{I}_{{C_i}}}\left( {{t_k}} \right) + {\bm{I}_{{C_j}}}\left( {{t_k}} \right)
\end{array}\label{merging_upd}
\end{equation}

\textbf{c) Out-dated CVT deleting :} After \textbf{a)} and \textbf{b)}, if there is a certain CVT with no VT observation for it, the CVT will be deleted after $t_d$ time slots.
\begin{equation}
  {\bm{C}_u}\left( {{t_k}} \right) = \left\{ \begin{array}{l}
\Phi ,{\ \ \ \ \ \ \ \ \ }\ \ {\rm if}\ \ \sum\limits_{t = {t_k} - {t_d}}^{{t_k}} {{\bm{I}_{{C_u}}}\left( t \right) = \bm{O}}  \vspace{0.5ex}\\
{\bm{C}_u}\left( {{t_k}} \right),{\ \ \ \ \rm{otherwise}}
\end{array} \right.\label{cvt_delete}
\end{equation}

As a summary, the CVT formation and cluster maintenance mechanism are shown in Algorithm 1.
Notice that ${\cal U}\left( {{t_k}} \right)$ in algorithm 1 is the motion information for each vehicle at time slot $t_k$, which will be introduced in (\ref{motion}$\sim$\ref{trans_define}) in next section.

\begin{figure*}[ht]
  \centering
  \includegraphics[width=1.8\columnwidth]{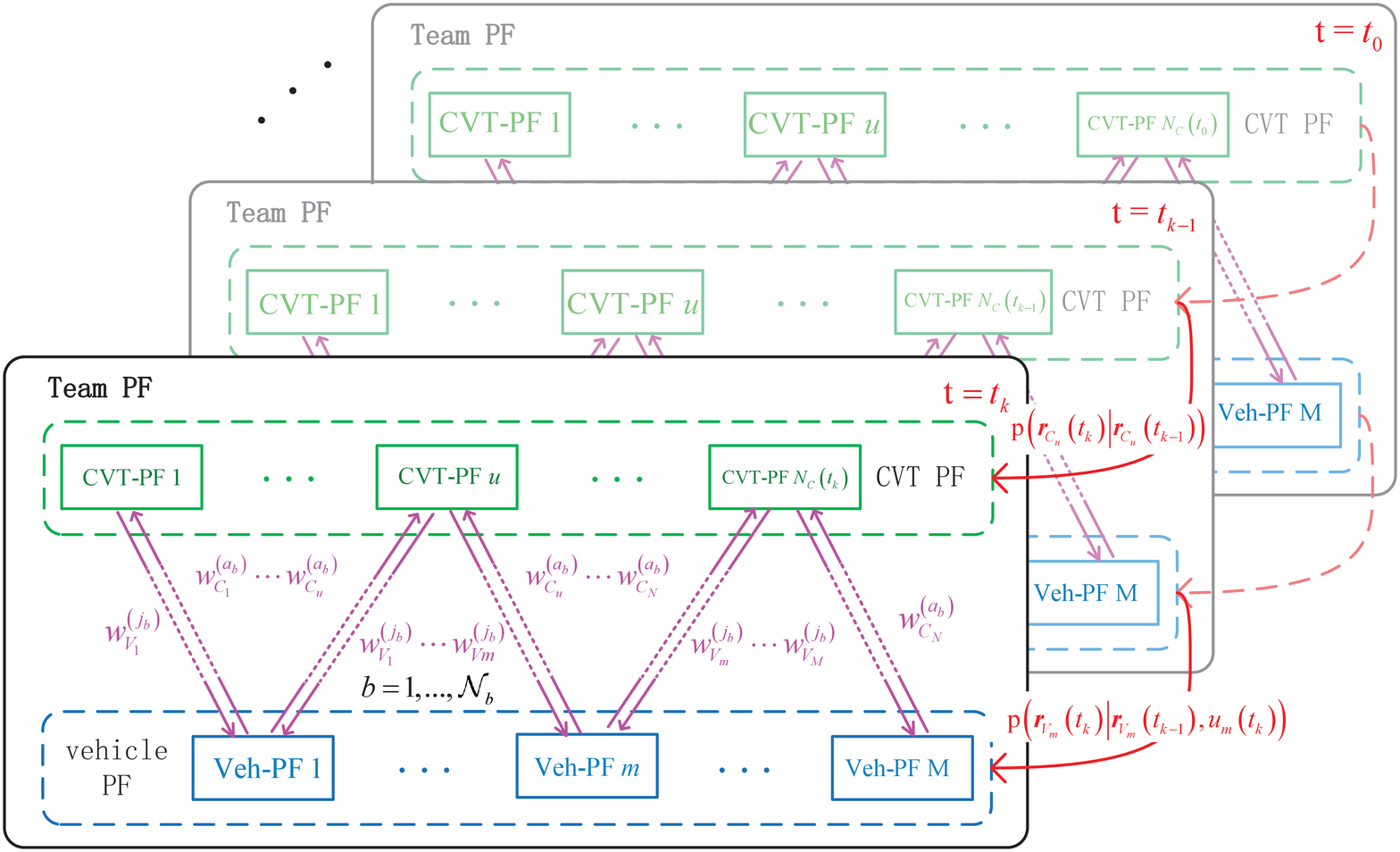}
  \caption{Team particle filter structure. For each time slot, \textbf{i)} both particle filters firstly update their particles from the previous time slot through ${\rm{p}}\left( {{\bm{r}_{{V_m}}}\left( {{t_k}} \right)\left| {{\bm{r}_{{V_m}}}\left( {{t_{k - 1}}} \right),{u_m}\left( {{t_k}} \right)} \right.} \right)$ and ${\rm{p}}\left( {{\bm{r}_{{C_u}}}\left( {{t_k}} \right)\left| {{\bm{r}_{{C_u}}}\left( {{t_{k - 1}}} \right)} \right.} \right)$. \textbf{ii)} both particle filters then update their particles iteratively by updating ${}^{\left( b \right)}w_{{V_m}}^{\left( j \right)}\left( {{t_k}} \right)$ and ${}^{\left( b \right)}w_{{C_u}}^{\left( f \right)}\left( {{t_k}} \right)$. \textbf{iii)} Finally the state of vehicles and CVTs are obtained by calculating ${\widehat {\bm{r}}_{{V_m}}}\left( {{t_k}} \right)$ and ${\widehat {\bm{r}}_{{C_u}}}\left( {{t_k}} \right)$.}
  \label{cpf}
\end{figure*}

\section{Multi-Vehicle localization and CVT Mapping}

In the above, we presented the mechanisms for CVT formation and maintenance. We now turn to the problem of vehicle localization, which is going to be performed jointly with that of CVT mapping. As this problem boils down to a large dimensional non-convex estimation problem, we propose a team particle filter based on particle filter \cite{van2001unscented,Siciliano2016Robotics,gustafsson2002particle}, which is well suited for this type of situation.
 As shown in Fig. \ref{cpf}, the team particle filters consist of CVT particle filters that correspond to possible 3D location candidates for the CVT, and vehicle particle filters that also correspond to possible 2D location candidates for the vehicles. For each time slot, both types of particles are updated as shown in Algorithm 2. At convergence, the probability density of CVT parameters and the location of vehicles are obtained.

The CVT particle filters and vehicle particle filters are introduced in section V-A and section V-B. The implementation of team particle filter is then introduced in section V-C.

\subsection{CVT Particle Filter}
In this subsection, ${N_C}\left( {{t_k}} \right)$ CVT particle filters are introduced to estimate the position of the ${N_C}\left( {{t_k}} \right)$ CVTs at the time slot $t_k$. In a CVT particle filter, the position of that CVT is approximated by ${\cal N}_C$ particles according to the Monte Carlo Sampling \cite{shapiro2003monte}, where each particle contains a 3-D position ${\bm{r}_{{C_u}}^{\left( a \right)}\left( {{t_k}} \right)}$ and a weight ${w_{{C_u}}^{\left( a \right)}\left( {{t_k}} \right)}$, $a = 1,...,{{\cal N}_C}$ for that 3-D position.
For example, ${\left[ {\bm{r}_{{C_u}}^{\left( a \right)}\left( {{t_k}} \right),w_{{C_u}}^{\left( a \right)}\left( {{t_k}} \right)} \right]}$ indicates the belief that CVT $u$ locates in ${\bm{r}_{{C_u}}^{\left( a \right)}\left( {{t_k}} \right)}$ is ${w_{{C_u}}^{\left( a \right)}\left( {{t_k}} \right)}$.
Hence the probability distribution function for the position of the $u$-th CVT can be approximated as:
\begin{equation}
  \begin{array}{*{20}{l}}
  {{\rm{p}}\left( {{\bm{r}_{{C_u}}}\left( {{t_k}} \right)\left| {{z_{\bm{p}_u^k}}\left( {{t_k}} \right)} \right.} \right)}\vspace{0.5ex}\\
  { \approx \sum\limits_{a = 1}^{{{\cal N}_C}} {w_{{C_u}}^{(a)}\left( {{t_k}} \right) \times \delta \left( {{\bm{r}_{{C_u}}}\left( {{t_k}} \right) - \bm{r}_{{C_u}}^{\left( a \right)}\left( {{t_k}} \right)} \right)} }
  \end{array}\label{CVT_pf}
\end{equation}
where we define the set of vehicles that observe a VT belonging to the CVT cluster $u$ at time $t_k$ as $\bm{V}_u^k$, and the set of its corresponding multi-paths as $\bm{p}_u^k$ in the $u$-th CVT.
\begin{equation}
  \begin{array}{l}
\bm{V}_u^k = \left\{ {m\left| {{\pi_{u,m}}\left( {{t_k}} \right) \ne 0} \right.} \right\}\vspace{0.5ex}\\
\bm{p}_u^k = \left\{ {\left( {m,{\pi_{u,m}}\left( {{t_k}} \right)} \right)\left| {m \in \bm{V}_u^k} \right.} \right\}
\end{array}
\end{equation}
Then the ${{z_{{\bm p}_u^k}}\left( {{t_k}} \right)}$ can be used to denote the multi-path observations from all the vehicles to the $u$-th CVT.

For the particle filter of the $u$-th CVT, according to \cite{van2001unscented,Siciliano2016Robotics,gustafsson2002particle,yin2018gnss}, the weight of its $a$-th particle is:

\begin{equation}
  \begin{array}{l}
w_{{C_u}}^{(a)}\left( {{t_k}} \right) \propto \frac{{{\rm{p}}\left( {{z_{{\bm p}_u^k}}\left( {{t_k}} \right)\left| {{\bm r}_{{C_u}}^{\left( a \right)}\left( {{t_k}} \right)} \right.} \right){\rm{p}}\left( {{\bm r}_{{C_u}}^{\left( a \right)}\left( {{t_k}} \right)\left| {{\bm r}_{{C_u}}^{\left( a \right)}\left( {{t_{k - 1}}} \right)} \right.} \right)}}{{{\rm{q}}\left( {{\bm r}_{{C_u}}^{\left( a \right)}\left( {{t_k}} \right)\left| {{\bm r}_{{C_u}}^{\left( a \right)}\left( {0:{t_{k - 1}}} \right),{z_{{\bm p}_u^k}}\left( {0:{t_k}} \right)} \right.} \right)}}\vspace{0.5ex}\\
 \ \ \ \ \ \ \ \ \ \ \ \times w_{{C_u}}^{(a)}\left( {{t_{k - 1}}} \right)
\end{array}\label{CVT_weight}
\end{equation}
where ${{\rm{q}}\left( {{\bm r}_{{C_u}}^{\left( a \right)}\left( {{t_k}} \right)\left| {{\bm r}_{{C_u}}^{\left( a \right)}\left( {0:{t_{k - 1}}} \right),{z_{{\bm p}_u^k}}\left( {0:{t_k}} \right)} \right.} \right)}$ is known as important distribution \cite{Siciliano2016Robotics}, which is set as CVT transition probability in this paper:
\begin{equation}
  \begin{array}{*{20}{l}}
{\rm{q}}\left( {\bm{r}_{{C_u}}^{\left( a \right)}\left( {{t_k}} \right)\left| {\bm{r}_{{C_u}}^{\left( a \right)}\left( {0:{t_{k - 1}}} \right),{z_{\bm{p}_u^k}}\left( {0:{t_k}} \right)} \right.} \right)\vspace{0.5ex}\\
  { = {\rm{p}}\left( {{\bm{r}_{{C_u}}^{\left( a \right)}}\left( {{t_k}} \right)\left| {{\bm{r}_{{C_u}}^{\left( a \right)}}\left( {{t_{k - 1}}} \right)} \right.} \right)}
  \end{array}\label{CVT_q_define}
\end{equation}
since the position of CVT is constant over time, so
\begin{equation}
{\rm{p}}\left( {{\bm{r}_{{C_u}}^{\left( a \right)}}\left( {{t_k}} \right)\left| {{\bm{r}_{{C_u}}^{\left( a \right)}}\left( {{t_{k - 1}}} \right)} \right.} \right){\rm{ = }}\delta \left( {{\bm{r}_{{C_u}}^{\left( a \right)}}\left( {{t_k}} \right) - {\bm{r}_{{C_u}}^{\left( a \right)}}\left( {{t_{k - 1}}} \right)} \right)\label{CVT_trans}
\end{equation}
Hence the CVT particle state can be drawn as
\begin{equation}
{\bm{r}_{{C_u}}^{\left( a \right)}}\left( {{t_k}} \right) \sim {\rm{p}}\left( {{\bm{r}_{{C_u}}^{\left( a \right)}}\left( {{t_k}} \right)\left| {{\bm{r}_{{C_u}}^{\left( a \right)}}\left( {{t_{k - 1}}} \right)} \right.} \right) \label{CVT_get_particle}
\end{equation}
and the weight updating equation is given by:
\begin{equation}
w_{{C_u}}^{(a)}\left( {{t_k}} \right) = w_{{C_u}}^{(a)}\left( {{t_{k - 1}}} \right){\rm{p}}\left( {{z_{\bm{p}_u^k}}\left( {{t_k}} \right)\left| {\bm{r}_{{C_u}}^{\left( a \right)}\left( {{t_k}} \right)} \right.} \right)\label{CVT_w_update}
\end{equation}
The prior in (\ref{CVT_w_update}) in team particle filter will be described in Section V-C in detail.

\subsection{Vehicle Particle Filter and Motion Update}
In this subsection, $M$ vehicle particle filters are introduced to estimate the position of the $M$ vehicles.
Similarly, in a vehicle particle filter, the position of that vehicle is approximated by ${\cal N}_V$ particles according to the Monte Carlo Sampling \cite{shapiro2003monte}, where each particle contains a 2-D position ${\bm{r}_{{V_m}}^{\left( j \right)}\left( {{t_k}} \right)}$ and a weight ${w_{{V_m}}^{\left( j \right)}\left( {{t_k}} \right)}$, $j = 1,...,{{\cal N}_V}$ for that 2-D position. For example, ${\left[ {\bm{r}_{{V_m}}^{\left( j \right)}\left( {{t_k}} \right),w_{{V_m}}^{\left( j \right)}\left( {{t_k}} \right)} \right]}$ indicates the belief that vehicle $m$ locates in ${\bm{r}_{{V_m}}^{\left( j \right)}\left( {{t_k}} \right)}$ is ${w_{{V_m}}^{\left( j \right)}\left( {{t_k}} \right)}$.
Hence the probability distribution function for the position of the $m$-th vehicle can be approximated as:
\begin{equation}
  \begin{array}{*{20}{l}}
{\rm{p}}\left( {{\bm{r}_{{V_m}}}\left( {{t_k}} \right)\left| {{z_m}\left( {{t_k}} \right),{u_m}\left( {{t_k}} \right)} \right.} \right)\vspace{0.5ex}\\
  { \approx \sum\limits_{j = 1}^{{{\cal N}_V}} {w_{{V_m}}^{(j)}\left( {{t_k}} \right)\delta \left( {{\bm{r}_{{V_m}}}\left( {{t_k}} \right) - \bm{r}_{{V_m}}^{\left( j \right)}\left( {{t_k}} \right)} \right)} }
  \end{array}\label{veh_pf}
\end{equation}

For the particle filter of the $m$-th vehicle, according to \cite{van2001unscented,Siciliano2016Robotics,gustafsson2002particle,yin2018gnss}, the weight of its $j$-th particle is:
\begin{equation}
  \begin{array}{l}
w_{{V_m}}^{(j)}\left( {{t_k}} \right) \propto \frac{{{\rm{p}}\left( {{\bm r}_{{V_m}}^{(j)}\left( {{t_k}} \right)\left| {{\bm r}_{{V_m}}^{(j)}\left( {{t_{k - 1}}} \right),} \right.{u_m}\left( {{t_k}} \right)} \right) \times {\rm{p}}\left( {{z_m}\left( {{t_k}} \right)\left| {{\bm r}_{{V_m}}^{(j)}\left( {{t_k}} \right)} \right.} \right)}}{{{\rm{q}}\left( {{\bm r}_{{V_m}}^{(j)}\left( {{t_k}} \right)\left| {{\bm r}_{{V_m}}^{(j)}\left( {{t_{k - 1}}} \right)} \right.,{z_m}\left( {{t_k}} \right),{u_m}\left( {{t_k}} \right)} \right)}}\vspace{0.5ex}\\
 \ \ \ \ \ \ \ \ \ \ \ \times w_{{V_m}}^{(j)}\left( {{t_{k - 1}}} \right)
\end{array}\label{veh_weight}
\end{equation}
set the vehicle transition probability as important distribution \cite{Siciliano2016Robotics,yin2018gnss}:
\begin{equation}
  \begin{array}{*{20}{c}}
  {{\rm{q}}\left( {\bm{r}_{{V_m}}^{(j)}\left( {{t_k}} \right)\left| {\bm{r}_{{V_m}}^{(j)}\left( {{t_{k - 1}}} \right)} \right.,{z_m}\left( {{t_k}} \right),{u_m}\left( {{t_k}} \right)} \right)}\vspace{0.5ex}\\
  { = {\rm{p}}\left( {\bm{r}_{{V_m}}^{(j)}\left( {{t_k}} \right)\left| {\bm{r}_{{V_m}}^{(j)}\left( {{t_{k - 1}}} \right)} \right.,{u_m}\left( {{t_k}} \right)} \right)}
  \end{array}\label{veh_q}
\end{equation}
so the state of vehicle particles can be drawn as:
\begin{equation}
    \bm{r}_{{V_m}}^{(j)}\left( {{t_k}} \right)\sim {\rm{p}}\left( {\bm{r}_{{V_m}}^{(j)}\left( {{t_k}} \right)\left| {\bm{r}_{{V_m}}^{(j)}\left( {{t_{k - 1}}} \right)} \right.,{u_m}\left( {{t_k}} \right)} \right) \label{veh_trans}
\end{equation}

In this work, we consider the case of noisy motion information to test the robustness of our method. Assume the motion information can be observed with Gaussian error:
\begin{equation}
  {\cal U}\left( {{t_k}} \right) = \left\{ {{u_1}\left( {{t_k}} \right),...,{u_m}\left( {{t_k}} \right),...,{u_M}\left( {{t_k}} \right)} \right\} \label{motion}
\end{equation}
where ${u_m}\left( {{t_k}} \right)$ is the motion for $m$-th vehicle,
\begin{equation}
  \begin{array}{*{20}{c}}
{u_m}\left( {{t_k}} \right) = \left\{ {{\bm{v}_{{V_m}}}\left( {{t_k}} \right)} \right\}\vspace{0.5ex}\\

{\bm{v}_{{V_m}}}\left( {{t_k}} \right){\rm{ = }}\bm{v}_{{V_m}}^{real}\left( {{t_k}} \right) + {n_v} \cdot {e^{j{n_\omega }}}
\end{array} \label{trans_define}
\end{equation}
where ${n_v}\sim {\rm{N}}\left( {0,\sigma _v^2} \right)$ and ${n_\omega}\sim{\rm{N}}\left( {0,\sigma _\omega^2} \right)$ is the noise of speed and orientation of speed that follow Gaussian distribution \cite{mendrzik2019enabling}.
The transmission model can be described as
\begin{equation}
  {\bm{r}_{{V_m}}}\left( {{t_k}} \right) = {\bm{r}_{{V_m}}}\left( {{t_{k - 1}}} \right) + \left( {{\bm{v}_{{V_m}}}\left( {{t_{k - 1}}} \right) + {\bm{v}_{{V_m}}}\left( {{t_k}} \right)} \right) \cdot \frac{T}{2}\label{trans_model}
\end{equation}
Then the weight update equation is given by:
\begin{equation}
w_{{V_m}}^{(j)}\left( {{t_k}} \right) = w_{{V_m}}^{(j)}\left( {{t_{k - 1}}} \right){\rm{p}}\left( {{z_m}\left( {{t_k}} \right)\left| {\bm{r}_{{V_m}}^{(j)}\left( {{t_k}} \right)} \right.} \right)\label{veh_w_update}
\end{equation}
The prior in (\ref{veh_w_update}) in team particle filter will be described in Section V-C in detail.

\begin{algorithm}[t]
\caption{Team Particle Filter}
\LinesNumbered
\KwIn{$vehicle\ particles$ : ${p^{\left( 0 \right)}}\left( {{\bm{r}_{{V_m}}}\left( {{t_k}} \right)} \right)$ \\
\setlength{\parindent}{3em}$CVT\ particles$ : ${p^{\left( 0 \right)}}\left( {{\bm{r}_{{C_u}}}\left( {{t_k}} \right)} \right)$
}
\KwOut{$estimated\ CVT\ state$ : $\bm{{\widehat{{\cal X}}}}\left( {{t_k}} \right)$\\
\setlength{\parindent}{4em}$estimated\ vehicle\ state$ : $\bm{{\widehat{{\cal C}}}}\left( {{t_k}} \right)$
}
\For{$b \le {{\cal N}_{b}}$}{
    $sub$-$process\ one\ :\ CVT\ mapping$\\
    \For{$u = 1:{N_C}\left( {{t_k}} \right)$}{
        Select CVT particle batch ${p^{\left( b \right)}}\left( {{\bm{r}_{{C_u}}}\left( {{t_k}} \right)} \right)$ \\
        Update $w_{{C_u}}^{\left( {{a_b}} \right)}\left( {{t_k}} \right)$ as (\ref{CVT_w_update}),(\ref{model_ob}$\sim$\ref{eq_particle_level_cvt_upd}) \\
        Resampling CVT particles \\
    }
    $sub$-$process\ two\ :\ vehicle\ localization$\\
    \For{$m = 1:M$}{
        Select vehicle particle batch ${p^{\left( b \right)}}\left( {{\bm{r}_{{V_m}}}\left( {{t_k}} \right)} \right)$ \\
        Update $w_{{V_m}}^{\left( {{j_b}} \right)}\left( {{t_k}} \right)$ as (\ref{veh_w_update}),(\ref{model_ob_veh}$\sim$\ref{eq_particle_level_veh_upd})\\
        Resampling vehicle particles\\
    }
    $sub$-$process\ three\ :\  state\ estimation$\\
    Calculate the position of vehicles $\bm{{\widehat r}}_{{V_m}}^{\left( b \right)}\left( {{t_k}} \right)$ as (\ref{estimation_veh})\\
    Calculate the position of CVT $\bm{{\widehat r}}_{{C_u}}^{\left( b \right)}\left( {{t_k}} \right)$ as (\ref{estimation_CVT}) \\
    \If{$\left\| {\bm{{\widehat r}}_{{V_m}}^{\left( b \right)}\left( {{t_k}} \right) - \bm{{\widehat r}}_{{V_m}}^{\left( {b - 1} \right)}\left( {{t_k}} \right)} \right\| < \xi $}{
        $break$\;
    }
}
Calculate ${\bm{\widehat r}_{{C_u}}}\left( {{t_k}} \right)$ and ${\bm{{\widehat r}}_{V_m}}\left( {{t_k}} \right)$ as (\ref{estimation_CVT}$\sim$\ref{estimation_veh})\\
\label{A_2}
\end{algorithm}

\subsection{Team Particle Filter Implementation}
Different from the classical particle filter reported in the literature \cite{van2001unscented,Siciliano2016Robotics,gustafsson2002particle,yin2018gnss}, the team particle filter combines the CVT particle filters and the vehicle particle filters to estimate the state of CVTs and vehicles jointly.
As shown in Algorithm 2, the team particle filter introduces a stochastic batch iteration method that divides the update process into ${{\cal N}_{b}}$ iterations. In each iteration, a stochastic batch of particles are chosen from the vehicle particles and CVT particles for the joint estimation of CVTs and vehicles.
Each iteration contains three sub-processes : CVT mapping, vehicle localization and state estimation.

\textbf{a) CVT Mapping :}
In each iteration, a batch of particles are chosen from each CVT particle filter stochastically and the weight of the particles in the chosen batch are then updated by (\ref{CVT_w_update}).

The particles for the $u$-th CVT particle filter at time slot $t_k$  are drawn by (\ref{CVT_get_particle}) denoting as ${{p^{\left( 0 \right)}}\left( {{{\bm r}_{{C_u}}}\left( {{t_k}} \right)} \right)}$, and then a particle batch is stochastically chosen from ${{p^{\left( 0 \right)}}\left( {{{\bm r}_{{C_u}}}\left( {{t_k}} \right)} \right)}$ in $b$-th iteration denoting as  ${p^{\left( b \right)}}\left( {{\bm{r}_{{C_u}}}\left( {{t_k}} \right)} \right),b=1,2,...,{\cal N}_b$. We then denote the particles contained in ${p^{\left( b \right)}}\left( {{\bm{r}_{{C_u}}}\left( {{t_k}} \right)} \right)$ as ${p^{\left( b \right)}}\left( {{{\bm r}_{{C_u}}}\left( {{t_k}} \right)} \right) = \left\{ {{\bm r}_{{C_u}}^{\left( {{a_b}} \right)}\left( {{t_k}} \right),w_{{C_u}}^{\left( {{a_b}} \right)}\left( {{t_k}} \right)} \right\}$,
thus the observation PDF in (\ref{CVT_w_update}) for the particle batch chosen in $b$-th iteration is given by:
\begin{equation}
  \begin{array}{l}
  {\rm{p}}\left( {{z_{{\bm p}_u^k}}\left( {{t_k}} \right)\left| {{\bm r}_{{C_u}}^{\left( {{a_b}} \right)}\left( {{t_k}} \right)} \right.} \right)\vspace{0.5ex}\\
   = \prod\limits_{{\bm p} = \left( {m,{p_m}} \right) \in {\bm p}_u^k} {\int {{\rm p}\left( {{z_{\bm p}}\left( {{t_k}} \right)\left| {{\bm r}_{{C_u}}^{\left( {{a_b}} \right)}\left( {{t_k}} \right),{{\bm r}_{{V_m}}}\left( {{t_k}} \right)} \right.} \right)} } \vspace{0.5ex}\\
   \ \ \ \ \ \ \ \ \ \ \ \ \ \ \ \ \ \ \ \ \times {\rm  p}\left( {{{\bm r}_{{V_m}}}\left( {{t_k}} \right)} \right){\rm{d}}{{\bm r}_{{V_m}}}\left( {{t_k}} \right)
  \end{array}\label{model_ob}
\end{equation}
where the update process in (\ref{model_ob}) is based on the sum effect of the multi-paths in ${\bm p}_u^k$.

We suppose the distance between the particle ${{\bm r}_{{C_u}}^{\left( {{a_b}} \right)}\left( {{t_k}} \right)}$ and the particle level estimated CVT $\bm{{\widehat r}}_{{C_u}}^{\left( {{\bm p},j} \right)}\left( {{t_k}} \right)$ follows Gaussian distribution, the conditional PDF in the right part of equation (\ref{model_ob}) can then be calculated.
\begin{equation}
  \bm{{\widehat r}}_{{C_u}}^{\left( {{\bm p},j} \right)}\left( {{t_k}} \right) = {\bm r}_{{V_m}}^{\left( j \right)}\left( {{t_k}} \right) + \mathord{\buildrel{\lower3pt\hbox{$\scriptscriptstyle\rightharpoonup$}}
\over R} \left( {{z_{\bm p}}\left( {{t_k}} \right)} \right),{\bm p} = \left( {m,{p_m}} \right) \in {\bm p}_u^k \label{eq_particle_level_cvt_upd}
\end{equation}
where $\bm{{\widehat r}}_{{C_u}}^{\left( {{\bm p},j} \right)}\left( {{t_k}} \right)$ denote the particle level estimation for the $u$-th CVT based on the observation from the multi-path in ${\bm p}_u^k$ and the $j$-th particle from the corresponding vehicle particle filter.

\textbf{b) Vehicle Localization :}
In each iteration, a batch of particles are chosen from each vehicle particle filter stochastically and the weight of the particles in the chosen batch are then updated by (\ref{veh_w_update}).

The particles for the $m$-th vehicle particle filter at time slot $t_k$ are drawn by (\ref{veh_trans}) denoting as ${{p^{\left( 0 \right)}}\left( {{{\bm r}_{{V_m}}}\left( {{t_k}} \right)} \right)}$, and then a particle batch is stochastically chosen from ${{p^{\left( 0 \right)}}\left( {{{\bm r}_{{V_m}}}\left( {{t_k}} \right)} \right)}$ in $b$-th iteration denoting as ${p^{\left( b \right)}}\left( {{\bm{r}_{{V_m}}}\left( {{t_k}} \right)} \right),b=1,2,...,{\cal N}_b$.
We then denote the particles contained in ${p^{\left( b \right)}}\left( {{\bm{r}_{{V_m}}}\left( {{t_k}} \right)} \right)$ as ${p^{\left( b \right)}}\left( {{{\bm r}_{{V_m}}}\left( {{t_k}} \right)} \right) = \left\{ {{\bm r}_{{V_m}}^{\left( {{j_b}} \right)}\left( {{t_k}} \right),w_{{V_m}}^{\left( {{j_b}} \right)}\left( {{t_k}} \right)} \right\}$, thus the observation PDF in (\ref{veh_w_update}) for the particle batch chosen in $b$-th iteration is given by:
\begin{equation}
  \begin{array}{l}
  {\rm{p}}\left( {{z_m}\left( {{t_k}} \right)\left| {{\bm r}_{{V_m}}^{({j_b})}\left( {{t_k}} \right)} \right.} \right)\vspace{0.5ex}\\
   = \prod\limits_{\begin{subarray}{c}
      {p_m} = 1,\\
      {{\pi _{{u_m} ,m}}\left( {{t_k}} \right) = {p_m}}
   \end{subarray}}^{{N_m}\left( {{t_k}} \right)} {\int {{\rm{p}}\left( {{z_{\left( {m,{p_m}} \right)}}\left( {{t_k}} \right)\left| {{\bm r}_{{V_m}}^{({j_b})}\left( {{t_k}} \right),{{\bm r}_{{C_{u_m} }}}\left( {{t_k}} \right)} \right.} \right)} } \vspace{0.5ex}\\
   \ \ \ \ \ \ \ \ \ \ \ \ \ \ \ \times {\rm{p}}\left( {{{\bm r}_{{C_{u_m} }}}\left( {{t_k}} \right)} \right){\rm{d}}{{\bm r}_{{C_{u_m} }}}\left( {{t_k}} \right)
  \end{array}\label{model_ob_veh}
\end{equation}
where the update process in (\ref{model_ob_veh}) is based on the sum effect of the multi-paths observed by the $m$-th vehicle, and $u_m$ is the index of CVT that observed by vehicle $m$ through its $p_m$-th multi-path at time slot $t_k$ satisfying ${{\pi _{{u_m} ,m}}\left( {{t_k}} \right) = {p_m}}$.

We suppose the distance between the particle ${{\bm r}_{{V_m}}^{({j_b})}\left( {{t_k}} \right)}$ and the particle level estimated vehicle position $\bm{{\widehat r}}_{{V_m}}^{\left( {{p_m},a} \right)}\left( {{t_k}} \right)$ follows Gaussian distribution, the conditional PDF in the right part of equation (\ref{model_ob_veh}) can then be calculated.
\begin{equation}
  \begin{array}{l}
\bm{{\widehat r}}_{{V_m}}^{\left( {{p_m},a} \right)}\left( {{t_k}} \right) = {\bm r}_{{C_{u_m} }}^{\left( a \right)}\left( {{t_k}} \right) - \mathord{\buildrel{\lower3pt\hbox{$\scriptscriptstyle\rightharpoonup$}}
\over R} \left( {{z_{\left( {m,{p_m}} \right)}}\left( {{t_k}} \right)} \right)\vspace{0.5ex}\\
{p_m} = 1,2,...,{N_m}\left( {{t_k}} \right),{\pi _{{u_m} ,m}}\left( {{t_k}} \right) = {p_m}
\end{array} \label{eq_particle_level_veh_upd}
\end{equation}
where $\bm{{\widehat r}}_{{V_m}}^{\left( {{p_m},a} \right)}\left( {{t_k}} \right)$ denote the particle level estimation for the $m$-th vehicle based on its $p_m$-th multi-path observations and the $a$-th particle from the corresponding CVT particle filter.

\textbf{c) State Estimation :} According to (\ref{CVT_pf}), the position of the $u$-th CVT is estimated as (time dimension is omitted for short):
\begin{equation}
  \begin{array}{*{20}{l}}
  {{\bm{{\widehat r}}_{C_u}}\left( {{t_k}} \right) = \int {{{\bm{r}}_{{C_u}}} \times {\rm{p}}\left( {{{\bm{r}}_{C_u}}\left| {{z_{{\bm{p}}_u^k}}} \right.} \right)} d{{\bm{r}}_{{C_u}}}}\\
  { \approx \int {{{\bm{r}}_{{C_u}}} \times \left( {\sum\limits_{a = 1}^{{{\cal N}_C}} {w_{{C_u}}^{(a)} \times \delta \left( {{{\bm{r}}_{{C_u}}} - {\bm{r}_{{C_u}}^{\left( a \right)}}} \right)} } \right)} d{{\bm{r}}_{{C_u}}}}\\
  { = \sum\limits_{a = 1}^{{{\cal N}_C}} {w_{{C_u}}^{(a)}}  \times \left( {\int {{{\bm{r}}_{{C_u}}} \times \delta \left( {{{\bm{r}}_{{C_u}}} - {\bm{r}_{{C_u}}^{\left( a \right)}}} \right)d{{\bm{r}}_{{C_u}}}} } \right)}\\
  { = \sum\limits_{a = 1}^{{{\cal N}_C}} {w_{{C_u}}^{(a)}\left( {{t_k}} \right) \times {\bm{r}_{{C_u}}^{\left( a \right)}}\left( {{t_k}} \right)} }
\end{array}\label{estimation_CVT}
\end{equation}

According to (\ref{veh_pf}), the position of the $m$-th vehicle is estimated as:
\begin{equation}
  \begin{array}{*{20}{l}}
  {{\bm{{\widehat r}}_{{V_m}}}\left( {{t_k}} \right) = \int {{\bm{r}_{{V_m}}} \times {\rm{p}}\left( {{\bm{r}_{{V_m}}}\left| {{z_m},{u_m}} \right.} \right)d{\bm{r}_{{V_m}}}} }\\
  { \approx \int {{\bm{r}_{{V_m}}} \times \sum\limits_{j = 1}^{{{\cal N}_V}} {w_{{V_m}}^{(j)} \times \delta \left( {{\bm{r}_{{V_m}}} - \bm{r}_{{V_m}}^{(j)}} \right)} d{\bm{r}_{{V_m}}}} }\\
  { = \sum\limits_{j = 1}^{{{\cal N}_V}} {w_{{V_m}}^{(j)} \times \left( {\int {{\bm{r}_{{V_m}}} \times \delta \left( {{\bm{r}_{{V_m}}} - \bm{r}_{{V_m}}^{(j)}} \right)d{\bm{r}_{{V_m}}}} } \right)} }\\
  { = \sum\limits_{j = 1}^{{{\cal N}_V}} {w_{{V_m}}^{(j)}\left( {{t_k}} \right) \times \bm{r}_{{V_m}}^{(j)}\left( {{t_k}} \right)} }
\end{array}\label{estimation_veh}
\end{equation}

As a summary of Section IV and V, the process of Team Channel-SLAM is shown in Algorithm 3.

\begin{algorithm}[t]
\caption{Team Channel-SLAM}
\LinesNumbered
\KwIn{$initial\ vehicle\ state$ : $\bm{{\cal X}}\left( {{t_0}} \right)$\\
\setlength{\parindent}{3em}$observations$ : ${\cal Z}\left( {{t_{1:K}}} \right)$\\
\setlength{\parindent}{3em}$motion\ information$ : ${\cal U}\left( {{t_{1:K}}} \right)$\\
}
\KwOut{$estimated\ vehicle\ state$ : $\bm{{\widehat{{\cal X}}}}\left( {{t_{1:K}}} \right)$\\
\setlength{\parindent}{4em}$estimated\ CVT\ state$ : $\bm{{\widehat{{\cal C}}}}\left( {{t_{1:K}}} \right)$
}
Initialize the initial particles $\bm{r}_{{V_m}}^{(j)}\left( {{t_0}} \right)$ from $\bm{{\cal X}}\left( {{t_0}} \right)$\\
\For{$t = {t_1}:{t_K}$}{
    $CVT\ Formation\ and\ Cluster\ Maintenance$ as Algorithm 1\\
    Draw vehicle particles ${{p^{\left( 0 \right)}}\left( {{{\bm r}_{{V_m}}}\left( {{t_k}} \right)} \right)}$ as (\ref{veh_trans})\\
    Draw CVT particles ${{p^{\left( 0 \right)}}\left( {{{\bm r}_{{C_u}}}\left( {{t_k}} \right)} \right)}$ as (\ref{CVT_get_particle})\\
    $Team\ Particle\ Filter$ as Algorithm 2\\
}
\label{A_3}
\end{algorithm}

\begin{figure*}[t]
  \centering
  \includegraphics[width=2.0\columnwidth]{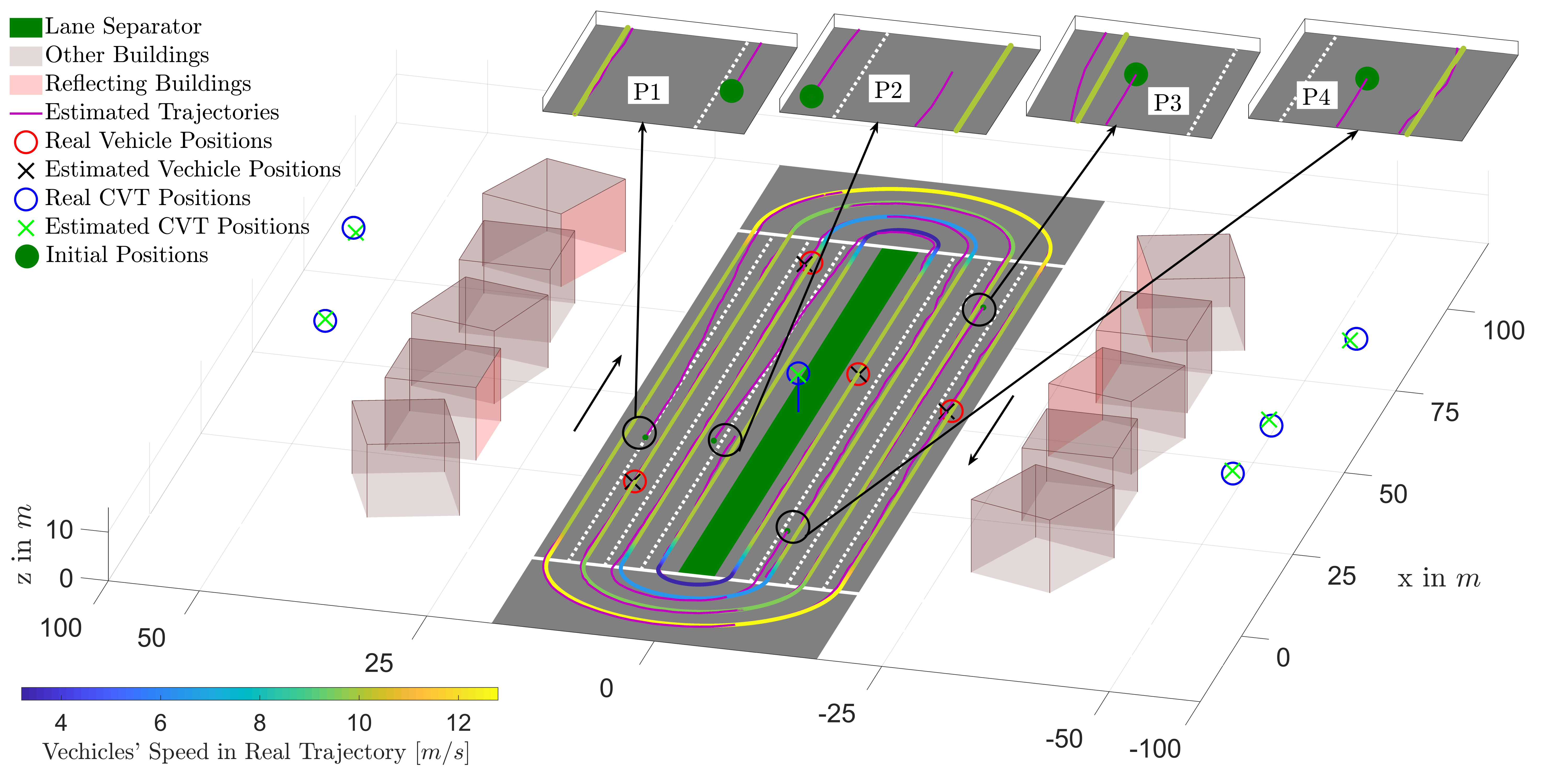}
  \caption{Simulation scenario and estimated trajectories. The estimated trajectories that the density of vehicles ${\rho _v} = 4\;\rm{vehicles}/(32 \times 132{\rm{m}^2})$ are shown by purple lines. The real trajectories and real speeds of the vehicles are indicated via the color bar. The estimated vehicle positions and CVT positions at the 150-th time slot are shown in the figure, and the reflecting planes at that time are indicated by the red-colored surface in the buildings. The four enlarged parts ("P1", "P2", "P3", and "P4") indicate the initial positions of the four vehicles.
  }
  \label{stp}
\end{figure*}

\section{Simulation Results}
In this section, a simulation experiment is carried out to test the performance of the proposed algorithm.
 Performance evaluation of ToA and AoA estimation in LTE was reported in \cite{blanco2019performance} and the results show that the median error of ToA and AoA is $e_d^{m}=1.76m$ and $e_\theta ^{m}={1.4^ \circ }$, respectively. Since the noise of ToA and AoA can be assumed as Gaussian distribution \cite{mendrzik2019enabling}, the variance of ToA and AoA error is calculated as:
\begin{equation}
  \begin{array}{*{20}{l}}
  {\int_{ - e_d^m}^{e_d^m} {\frac{1}{{\sqrt {2\pi } {\sigma _d}}}} {e^{ - \frac{{{x^2}}}{{2\sigma _d^2}}}}dx = \frac{{\rm{1}}}{{\rm{2}}} \Rightarrow {\sigma _d} = 2.61{\rm{m }}} \vspace{0.5ex}\\
  {\int_{ - e_\theta ^m}^{e_\theta ^m} {\frac{1}{{\sqrt {2\pi } {\sigma _\theta }}}} {e^{ - \frac{{{x^2}}}{{2\sigma _\theta ^2}}}}dx = \frac{{\rm{1}}}{{\rm{2}}} \Rightarrow {\sigma _\theta } = 2.08^\circ }
  \end{array}\label{variance}
\end{equation}
As a result, building on the recommendation of \cite{blanco2019performance}, the variance of the distance measurement noise ${\sigma _d}$ was set to $ 2.61{\rm{m}}$, and the variance of the angle measurement noise ${\sigma _\alpha }$ was set to ${2.08^ \circ }$ (both for polar angle and azimuth angle).
As for the noise of motion information, the variance of speed error and its orientation error set as is ${\sigma _v} = 0.1m/s$ and ${\sigma _\omega } = 0.1\deg /s$ \cite{mendrzik2019enabling}. In this paper, we suppose the initial position of vehicle is get from GPS, whose error follows a Gaussian distribution with $\sigma _r^{gps} = 3m$. Since there is infinity in Gaussian distribution, the errors mentioned above are all cut by $2\sigma $ to make it more reasonable.

The sampling interval $t_\delta$ is set as ${t_\delta } = 0.1{\rm{s}}$, the out-dated CVT cluster deleting time delay $t_d$ is set as ${t_d} = 10 * {t_\delta }$.
The data association threshold ${L_A}$ and CVT cluster merging threshold ${L_M}$ are set as ${L_A} = {L_M} = - 2.36$, which indicate the near worst VT estimation error from equation (\ref{cal_VTs}).
\begin{equation}
  \begin{array}{c}
  {L_A} = {L_M} =  - \ln \left( {{\varepsilon _{\mathop{\rm m}\nolimits} } + 1} \right) \vspace{0.5ex}\\
  \varepsilon _{\mathop{\rm m}\nolimits} ^2 = {\left( {{d_{\mathop{\rm m}\nolimits} } + n{\sigma _d}} \right)^2} + d_{\mathop{\rm m}\nolimits} ^{\rm{2}}{\rm{ - 2}}{d_{\mathop{\rm m}\nolimits} }\left( {{d_{\mathop{\rm m}\nolimits} } + n{\sigma _d}} \right)\cos \left( {n{\sigma _\theta }} \right)
  \end{array}\label{la_lm}
\end{equation}
where ${{\varepsilon _{\mathop{\rm m}\nolimits} }}$ is the near worst VT estimation error, ${{d_{\mathop{\rm m}\nolimits} }} = 100m$ is the maximum communication range for a base station and $n=2$, which means that the cumulative probability in Gaussian distribution is over 0.95 ($\Phi \left( {2\sigma } \right) = 0.9544$). So the setting of $L_A$ and $L_M$ indicates that the 95.44 percentile of the ToA and AoA error is considered within the data association threshold and CVT cluster merging threshold to cover a near worst situation.

Totally 300 time slots' movement of the vehicles are tracked under different vehicle densities (${t_K} = 300 * {t_\delta }$). The particle number for both CVT PF and vehicle PF is set to 120 (${{\cal N}_C} = {{\cal N}_V} = 120$), which shows a acceptable compromise between the accuracy and complexity. The number of iterations in Team Channel-SLAM ${\cal N}_b$ is set as ${\cal N}_b = 10$ to ensure that there are as much iterations as possible and also that there are enough particles in each particle batch according to the number of particles in vehicle particle filter and CVT particle filter. The results are based on 100 simulation runs.

\subsection{Vehicle Trajectory and Scenario Setting}
As shown in Fig. \ref{stp}, the base station locates at $\left[ {50m,0,8m} \right]$, and the buildings distribute besides the road. The position of the buildings are unknown in TCS, and the position of base station is also not required in this paper because authentication may be needed to get its position for security reasons. There are totally 8 lanes in the road, the width of each lane is 4 meters. The vehicles run clockwise along the four trajectories, and the speed of each vehicle is indicated by the color bar. Three kinds of motion models (uniform motion,  uniformly accelerated motion and uniform circular motion) are considered in the trajectory in order to make the simulation more realistic. All the vehicles drive within the area within $\left[ { - 16m,16m} \right]$ of $y$-axis and $\left[ { 0m,132m} \right]$ of $x$-axis. Since the number of vehicles will influence the performance of the algorithm, we explore the performance of TCS under different vehicle densities. We define the number of vehicles per unit area as vehicle density denoted as ${\rho _V}$. For example, the vehicle density in Fig. \ref{stp} is ${\rho _V} = 4\;\rm{vehicles}/(32 \times 132{\rm{m}^2}) \approx 9.47 \times {10^{ - 4}}\;\rm{vehicle}/{\rm{m}^2}$. For convenience,
  we use ${\rm{vehicle/}}\left( {{\rm{32}} \times {\rm{132}}{{\rm{m}}^2}} \right)$ as the dimension of vehicle density. We explore the performance of TCS when ${\rho _V} = 1,2,4,8,12,16,24\;{\rm{vehicle/}}\left( {{\rm{32}} \times {\rm{132}}{{\rm{m}}^2}} \right)$.

Fig. \ref{stp} also shows the estimated trajectory of vehicles when vehicle density ${\rho _V} = 4\;{\rm{vehicles/}}\left( {{\rm{32}} \times {\rm{132}}{{\rm{m}}^2}} \right)$. The initial position of the four vehicles are circled by black lines and also enlarged in "P1", "P2", "P3", and "P4". We can see that the initial positioning error of the vehicles are large and the estimated vehicle trajectories get gradually close to the real trajectory.
So we can see preliminary from Fig. \ref{stp} that the vehicles are well tracked and CVTs are well estimated by our algorithm. However a more precise quantitative analysis is done in the following subsections.

\begin{figure*}[t]
\centering
\subfigure[CVT particles converging to the real CVT position over time (the depth of the color represent the density of the particles).]{\includegraphics[width=18cm]{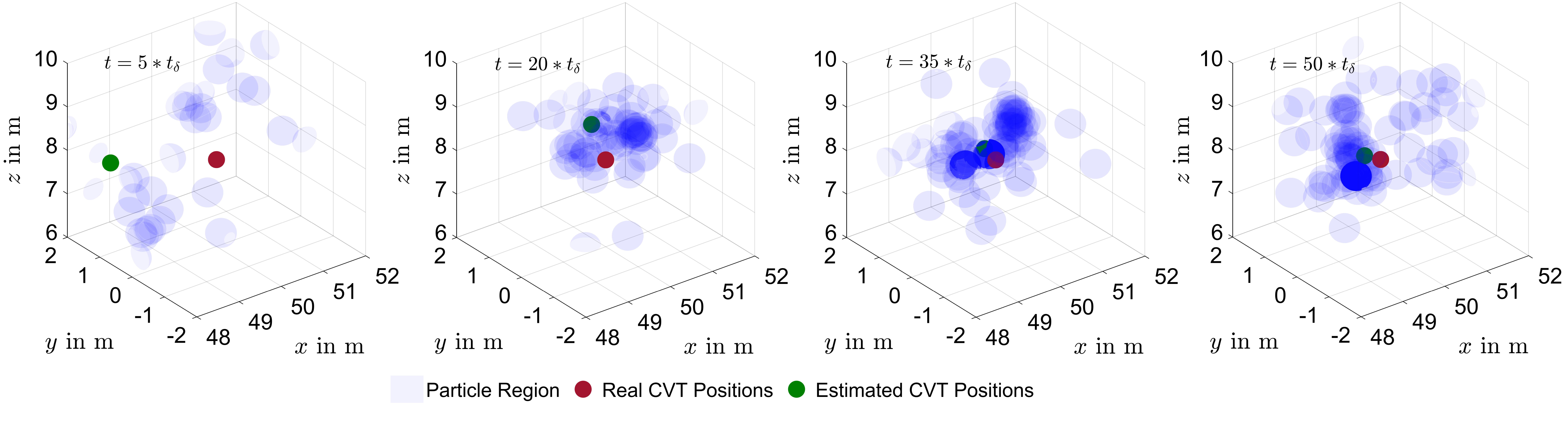}} \\
\subfigure[$80\%$ level vehicle positioning error $\varepsilon$ over time under different vehicle densities ${\rho _V} = 1,2,4,8,16,24\ {\rm{vehicle/}}\left( {{\rm{32}} \times {\rm{132}}{{\rm{m}}^2}} \right)$. ]{\includegraphics[width=9cm]{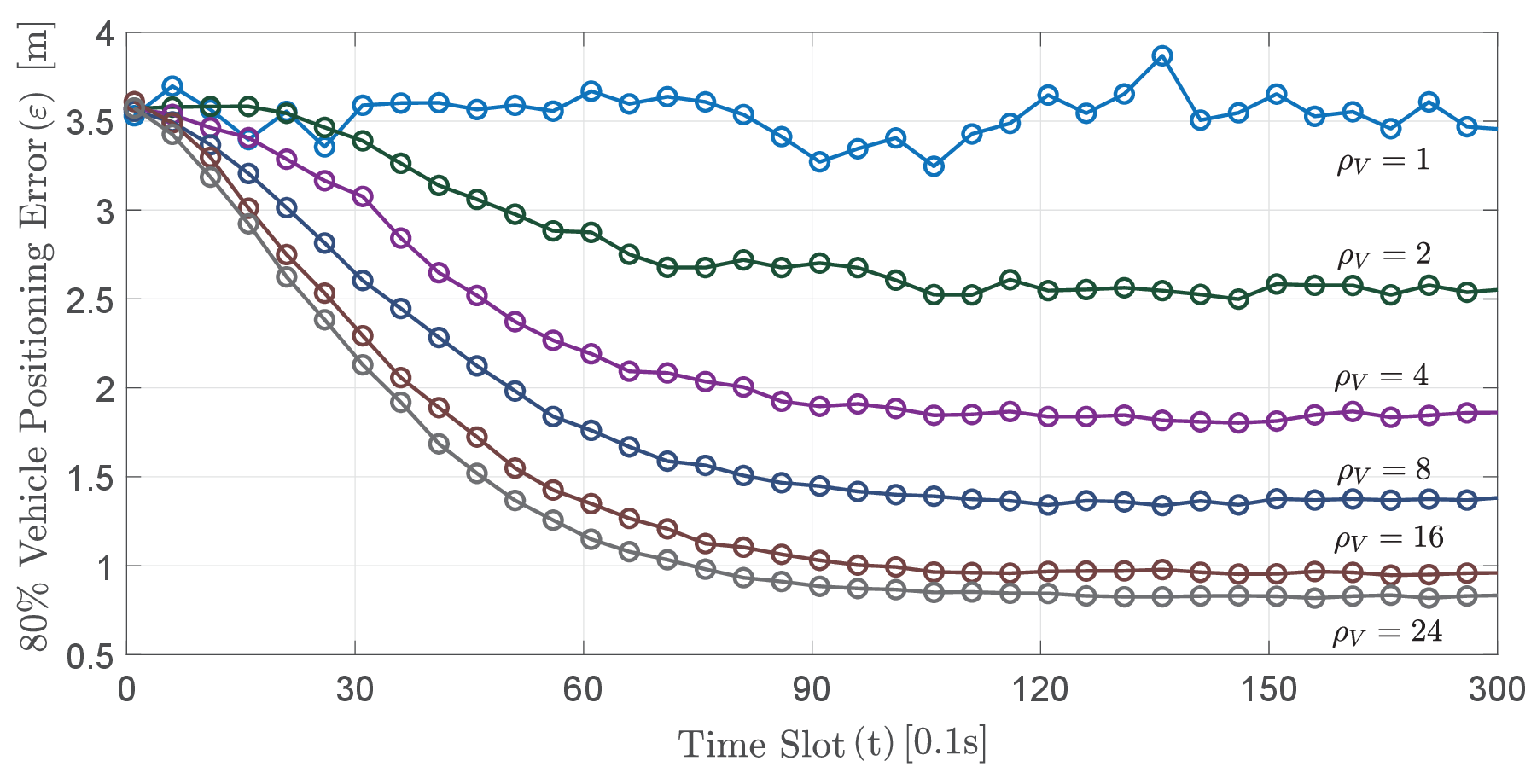}}
\subfigure[justification=centering][MAE Positioning errors of Team Channel-SLAM and Channel-SLAM over different vehicle densities. ]{\includegraphics[width=9cm]{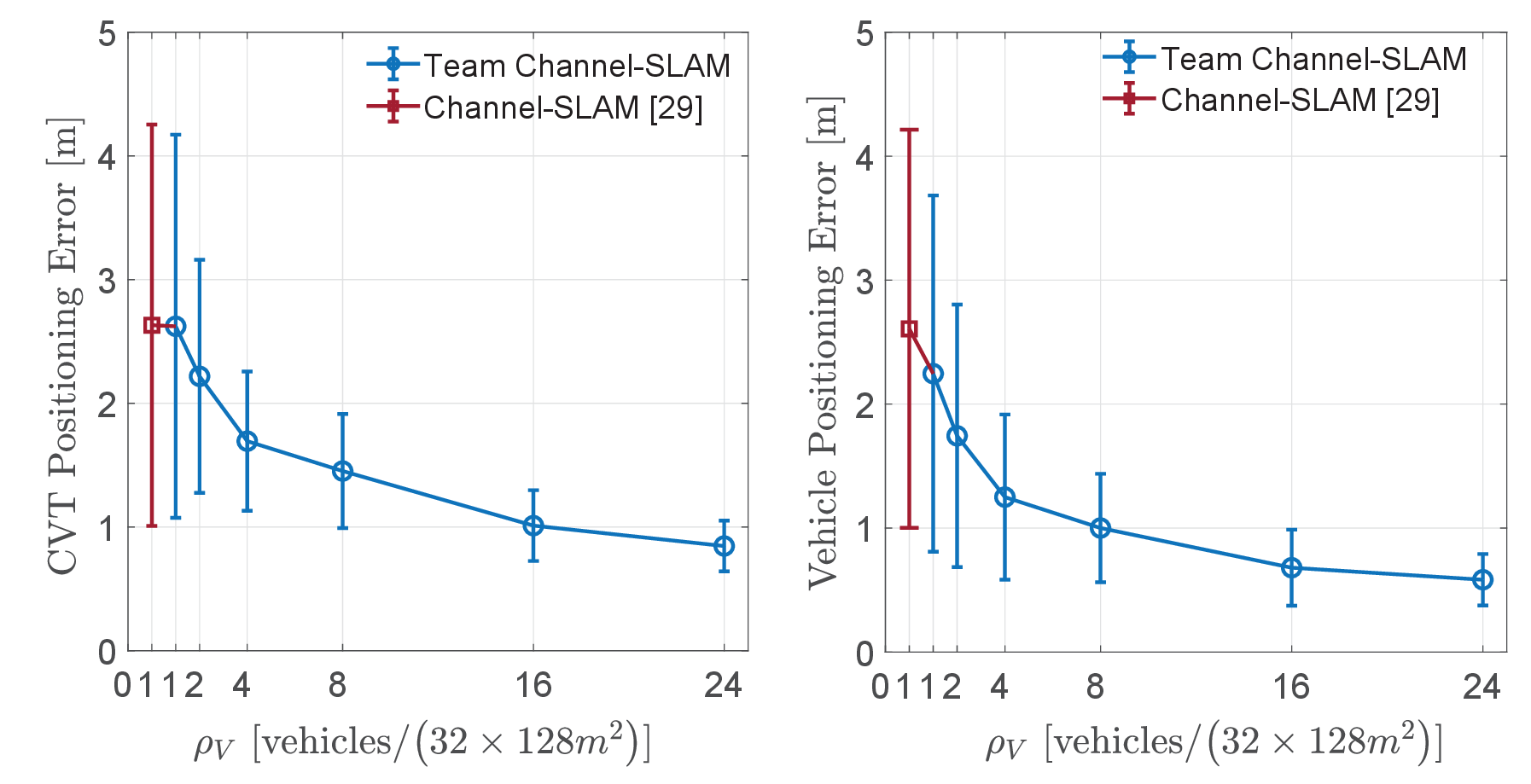}}%
\caption{Particle convergence of TPF and the positioning errors over time and different vehicle densities.
Note that in (a) only particles locating in $x \in \left[ {48,52} \right],y \in \left[ { - 2,2} \right]$ and $z \in \left[ {6,10} \right]$ are shown. (b) shows the 80\% level vehicle positioning error [$\varepsilon \left| {p\left( {{\varepsilon _V} \le \varepsilon } \right) = 0.8} \right.$] over time under different vehicle densities $\left[ {{\rm{vehicle/}}\left( {{\rm{32}} \times {\rm{132}}{{\rm{m}}^2}} \right)} \right]$. (c) shows the MAE (Mean Absolute Error) positioning errors for vehicles and CVT $\left[ {{e_V},{e_C}\left| {{e_V} = \mathbb{E}\left( {\left| {{\varepsilon _V}} \right|} \right),{e_C} = \mathbb{E}\left( {\left| {{\varepsilon _C}} \right|} \right)} \right.} \right]$ utilizing the Team Channel-SLAM algorithm and Channel-SLAM algorithm over different vehicle densities $\left[ {{\rm{vehicle/}}\left( {{\rm{32}} \times {\rm{132}}{{\rm{m}}^2}} \right)} \right]$.
}
\label{fig_particle_clustering}
\end{figure*}

\begin{figure*}[ht]
\centering
\subfigure[Cumulative probability density of CVT positioning error ]{\includegraphics[width=9cm]{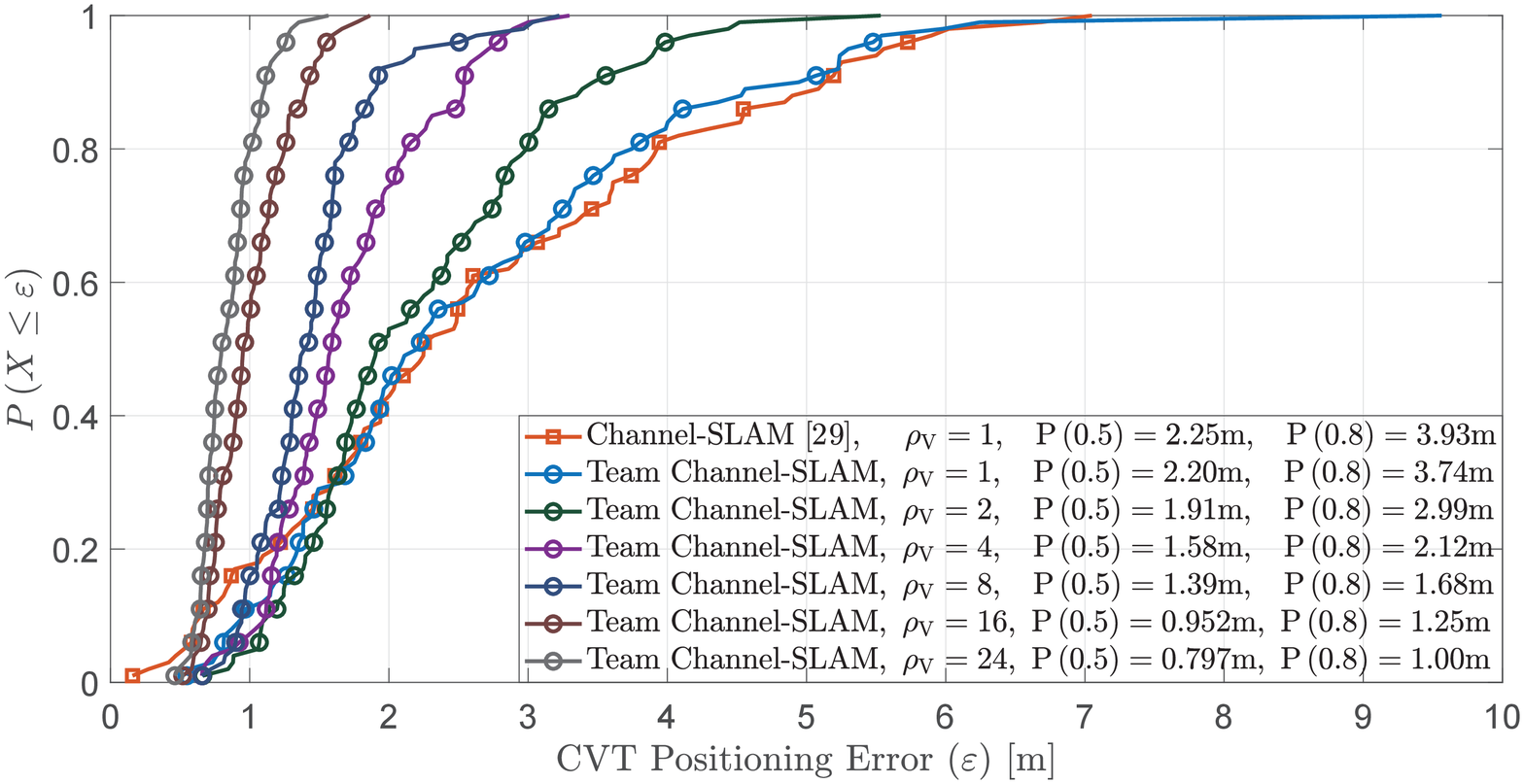}}%
\subfigure[Cumulative probability density of vehicle positioning error ]{\includegraphics[width=9cm]{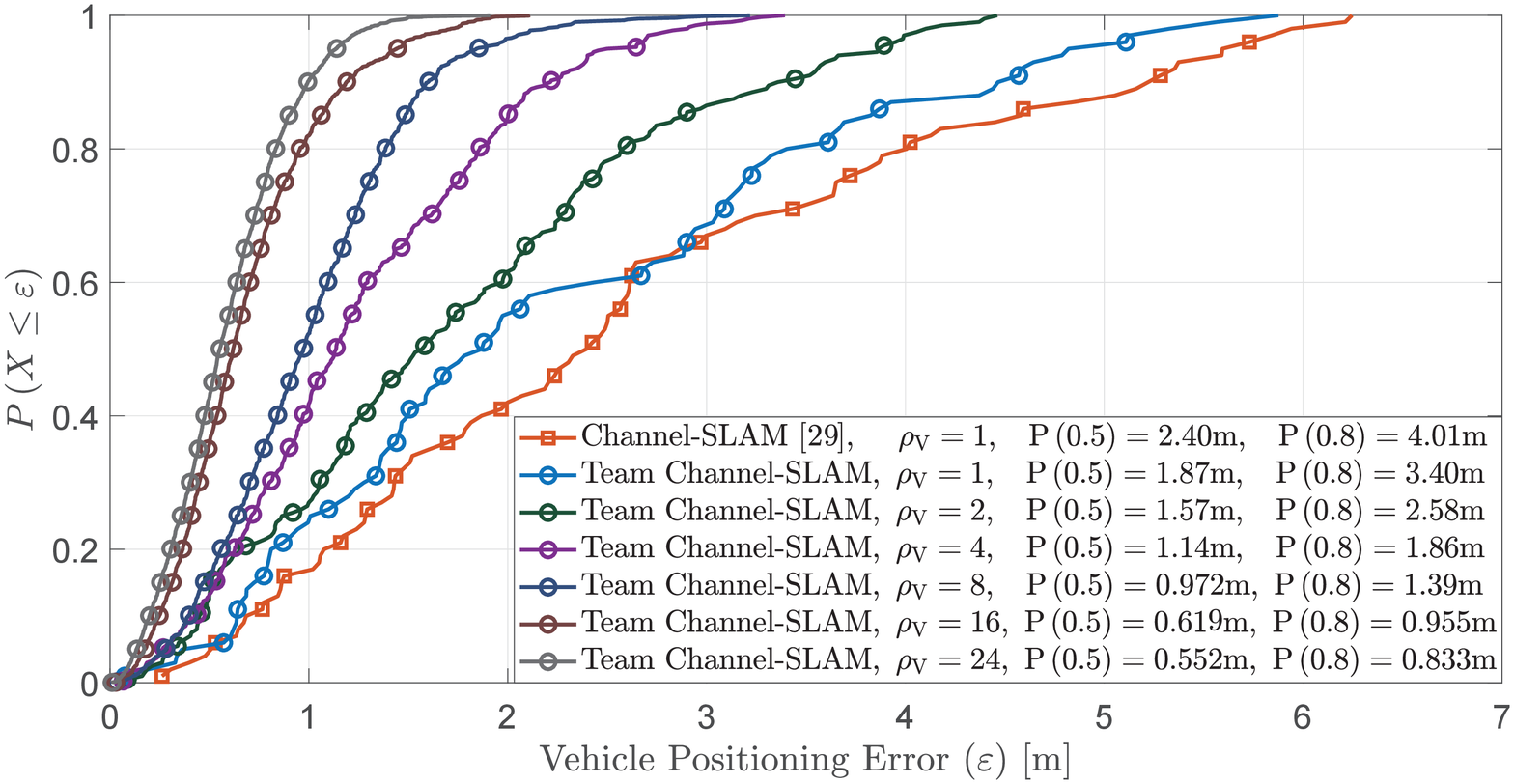}}%
\caption{Cumulative distribution function (CDF) of vehicle positioning error and CVT positioning error. (a) shows the CDF of vehicle positioning error over different vehicle densities $\left[ {{\rm{vehicles/}}\left( {{\rm{32}} \times {\rm{132}}{{\rm{m}}^2}} \right)} \right]$ for both TCS and CS \cite{gentner2016multipath}.  (b) shows the CDF of CVT positioning error over different vehicle densities for both TCS and CS \cite{gentner2016multipath}.}
\label{fig_cdf_CVT_veh}
\end{figure*}

\subsection{Particle Convergence}
Fig. \ref{fig_particle_clustering} (a) shows the particles for a certain CVT at \textbf{a)} $t = 5*{t_\delta }$, \textbf{b)} $t = 20*{t_\delta }$, \textbf{c)} $t = 35*{t_\delta }$ and \textbf{d)} $t = 50*{t_\delta }$. The position of the CVT is indicated by a red circle and what needs to be explained is that in order to show the distribution of particles more clearly, the particles are indicated by semitransparent blue balls. Thereby the area with higher depth of color indicates the higher possibility that the CVT locates here. From Fig. \ref{fig_particle_clustering} (a), we can see that the particles get closer to the true position of the CVT from $t = 5*{t_\delta}$ to $t = 50*{t_\delta}$. This indicates that CVT formation and cluster maintenance mechanism can associate and maintain the CVT clusters well so that the TPF can converge the CVT particles well over time.

\subsection{Error Analysis}

In order to show the performance of TCS, Channel-SLAM \cite{gentner2016multipath,Casella1996Rao} is introduced as a benchmark in this paper. Based on RBPF (Rao-Blackwellized particle filter) \cite{Casella1996Rao}, Channel-SLAM utilizes the super-particle filters (super-PF) and the sub-particle filters (sub-PF) to estimate the state of a single-vehicle and its corresponding VTs.
In Channel-SLAM, the super-PF is used for vehicle positioning and the sub-PF is used for VT positioning.
We set the particle number of super-PF the same as vehicle PF in TCS (${{\cal N}_{\rm{super}}} = {{\cal N}_{\rm{V}}} = 120$) and the particle number of sub-PF the same as CVT PF in TCS (${{\cal N}_{\rm{sub}}}={{\cal N}_C}=120$).
However, in Channel-SLAM, each particle in super-PF possess ${{N}_m}\left( {{t_k}} \right)$ sub-PFs,
so the total particle number of sub-PF (the particles used for VT estimation) is ${{\cal N}_{{\rm{sub}}}} \times {{\cal N}_{\rm{super}}} \times {{ N}_m}\left( {{t_k}} \right)$ in Channel-SLAM, compared with ${{\cal N}_C} \times {{ N}_m}\left( {{t_k}} \right)$ in TCS.

\subsubsection{Positioning Error over Vehicle Density}

Fig. \ref{fig_particle_clustering} (c) shows the mean absolute positioning error of vehicles $e_V$ and CVTs $e_C$ over different vehicle densities, which are defined as:
\begin{equation}
  \left[ {{e_V},{e_C}\left| {{e_V} = \mathbb{E}\left( {\left| {{\varepsilon _V}} \right|} \right),{e_C} = \mathbb{E}\left( {\left| {{\varepsilon _C}} \right|} \right)} \right.} \right] \label{eq_mae_error_def}
\end{equation}
where ${\varepsilon _V}$ and ${\varepsilon _C}$ denote the vehicle positioning errors and CVT positioning errors in the 100 simulation runs, respectively.

We can see that the positioning error and its variance have an obvious decreasing tendency when vehicle density increases. Specifically, when there are 4 vehicles among the $32 \times 132$ square meters' road, the positioning error of the vehicle has decreased $44.35\% $ than single-vehicle scenario.

This is because the increasing number of vehicles brings more multi-path observations for CVTs, which improves CVT positioning accuracy in the sub-process one of TPF in Algorithm 2. Then this improvement will provide better estimation for vehicle localization in the sub-process two of TPF in Algorithm 2. Thus a positive feedback is constituted between the two sub-processes in TPF to improve the position of vehicles and CVTs.
What's more, higher precision of CVTs
 and vehicles will make newly observed VTs more precisely associated into their corresponding CVT clusters in Algorithm 1, which also
 improves the accuracy of the CVT formation and cluster maintenance process in section IV. Thus another positive feedback is then constituted between TPF and CVT formation and cluster maintenance mechanism to futher improve the precision of multi-vehicle localization and CVT mapping.

\subsubsection{Positioning Error over Time}

Fig. \ref{fig_particle_clustering} (b) show the $80\%$ confidence level of vehicle positioning error over time under different vehicle densities,
\begin{equation}
{\rm p}\left( {{\varepsilon _V} \le \varepsilon } \right) = 0.8 \label{0_8_error}
\end{equation}
which indicates that for each value $\varepsilon$ in the figure, there is $80\%$ confidence that the vehicle positioning error ${\varepsilon _V}$ is no bigger than $\varepsilon$.

We can see from Fig. \ref{fig_particle_clustering} (b) that the positioning error of vehicle decreases over time and finally converge to a low value, which means that TCS can help the vehicle particles converge to their real position.
The reason behind this is that TCS will extract more and more information from added observations to make the CVT estimations get closer to their real positions gradually, which in return
helps to improve the accuracy of vehicle positioning as described in Section VI-C-1).

\subsubsection{CDF of Positioning Error}

Fig. \ref{fig_cdf_CVT_veh} (a) and Fig. \ref{fig_cdf_CVT_veh} (b) show the cumulative probability density (CDF) of the positioning error of vehicles and CVTs. Both $50\%$ and $80\%$ confidence level errors are labeled in the figure.

\subsubsection{Team Channel-SLAM and Channel-SLAM}
Seen from Fig. \ref{fig_particle_clustering} (c), the VT positioning error of Channel-SLAM is almost the same as TCS when ${\rho _V} = 1\;\rm{vehicle}/\left( {32 \times 132{\rm{m}^2}} \right)$, however, the vehicle positioning error of Channel-SLAM is higher than TCS when ${\rho _V} = 1\;\rm{vehicle}/\left( {32 \times 132{\rm{m}^2}} \right)$, which means that TCS have a better performance in particle convergence due to the CVT formation and cluster maintenance mechanism.

Additionally, we can found that the positioning error doesn't decrease over time when ${\rho _V} = 1\;\rm{vehicle}/\left( {32 \times 132{\rm{m}^2}} \right)$ in Fig. \ref{fig_particle_clustering} (b). This is because in this situation the CVT estimation is based on the relative observation (ToA and AoA) and the initial position of just one vehicle.
So the initial error of that vehicle will contaminate the CVT estimations for there is only one observation to each CVT.
 However, TCS is also useful in this situation, because it will eliminate the accumulative error from motion information and also provide a continuous and stable positioning for vehicles in case that the GPS or other positioning sensors get limited by the weather or other hard conditions.
However, with the increasing of the vehicle density, the CVTs get gradually better estimated, which results in higher precision of vehicle positioning.

\subsection{Performance over Building Density}

\begin{figure}[t]
  \centering
  \includegraphics[width=0.9\columnwidth]{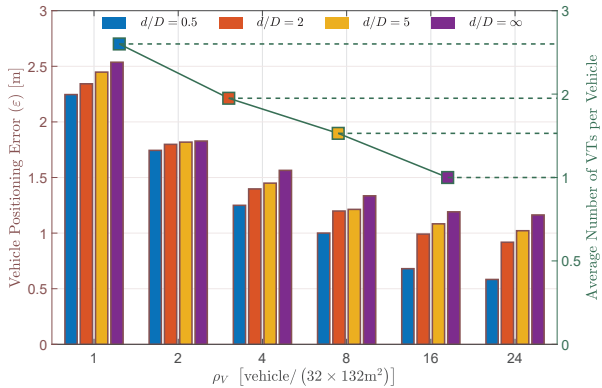}
  \caption{Vehicle positioning error and the average number of VTs observed by a vehicle at a time slot over different densities of buildings. The histogram shows the vehicle positioning error over different densities of buildings referring to the left axis, and the green line shows the average number of VTs over different densities of buildings referring to the right axis. Note that $d$ refers to the distance between buildings and $D$ refers to the length of buildings, so the larger the value $d/D$, the less dense of the buildings. }
  \label{fig_veh_err_density_ref}
\end{figure}

Fig. \ref{fig_veh_err_density_ref} shows the performance of vehicle positioning over different building densities. In this simulation, we set the length of each building constant as $D = 12{\rm m}$, and then change the distance between the buildings as $d=6{\rm m}, d=24{\rm m}, d=60{\rm m}$ as well as $d = \infty $ (indicates that there is no building). So the building density changes while the value $d$ changes, the value $d/D$ can then be used to indicate the density of the buildings. In detail, the larger the value, the less dense of the buildings.

The different densities of  buildings will then lead to different possibilities for a vehicle to observe a multi-path reflected by those buildings, and then influence the number of observed VTs. Thus we plot the average number of VTs observed by a vehicle at a time slot in the figure by green lines (referring to the right axis), which can be indicated that with the decreasing of the building density, the average number of the VTs observed by each vehicle decreases.

At the same time, we also test the performance of vehicle positioning under those different densities of buildings over different vehicle densities, where the result can be seen in the figure by the histogram (referring to the left axis). We can see that the denser the buildings, the more accurate of the vehicle positioning. Specifically, the positioning error when the building density $d/D=0.5$ is averagely 25.67\% better than the situation when the building density $d/D=\infty$ (indicates that there is no building). So it can be then indicated that having more virtual transmitters will lead to more information regarding the vehicle position and therefore improve localization accuracy, which is rather reasonable as the explanation in Section VI-C-1).

\section{Conclusion}
Team Channel-SLAM firstly establishes and maintains CVT clusters dynamically through CVT Formation and Cluster Maintenance mechanism and then estimates the position of multiple vehicles and CVTs simultaneously through Team Particle Filters. Simulation results show that TCS can improve the precision of vehicle localization for its better estimation for CVTs in multiple vehicle scenario. As from expectation, the vehicle positioning error keeps decreasing with the increasing density of vehicles and buildings in road traffic.

\section*{Acknowledgment}

This work is partially supported by Research on the key technology of seamless handover in cloud-RAN for network assisted automatic driving, NSFC project, under grant 61801047. It is also supported by the Beijing Nova Program of Science and Technology under grant Z191100001119028.

\bibliographystyle{IEEEtran}
\bibliography{IEEEabrv,reference}

\begin{thebibliography}{10}
\providecommand{\url}[1]{#1}
\csname url@samestyle\endcsname
\providecommand{\newblock}{\relax}
\providecommand{\bibinfo}[2]{#2}
\providecommand{\BIBentrySTDinterwordspacing}{\spaceskip=0pt\relax}
\providecommand{\BIBentryALTinterwordstretchfactor}{4}
\providecommand{\BIBentryALTinterwordspacing}{\spaceskip=\fontdimen2\font plus
\BIBentryALTinterwordstretchfactor\fontdimen3\font minus
  \fontdimen4\font\relax}
\providecommand{\BIBforeignlanguage}[2]{{%
\expandafter\ifx\csname l@#1\endcsname\relax
\typeout{** WARNING: IEEEtran.bst: No hyphenation pattern has been}%
\typeout{** loaded for the language `#1'. Using the pattern for}%
\typeout{** the default language instead.}%
\else
\language=\csname l@#1\endcsname
\fi
#2}}
\providecommand{\BIBdecl}{\relax}
\BIBdecl

\bibitem{9145339}
X.~{Chu}, Z.~{Lu}, L.~{Wang}, X.~{Wen}, and D.~{Gesbert}, ``Team channel-slam:
  A cooperative mapping approach to vehicle localization,'' in \emph{2020 IEEE
  International Conference on Communications Workshops (ICC Workshops)}, 2020,
  pp. 1--6.

\bibitem{DT}
D.~for Transport, ``Research on the impacts of connected and autonomous
  vehicles on traffic flow,'' \url {https://trid.trb.org/view/1448451}, 2016.

\bibitem{kuutti2018survey}
S.~Kuutti, S.~Fallah, K.~Katsaros, M.~Dianati, F.~Mccullough, and
  A.~Mouzakitis, ``A survey of the state-of-the-art localization techniques and
  their potentials for autonomous vehicle applications,'' \emph{IEEE Internet
  of Things Journal}, vol.~5, no.~2, pp. 829--846, 2018.

\bibitem{ulmschneider2016multipath}
M.~Ulmschneider, R.~Raulefs, C.~Gentner, and M.~Walter, ``Multipath assisted
  positioning in vehicular applications,'' in \emph{2016 13th Workshop on
  Positioning, Navigation and Communications (WPNC)}.\hskip 1em plus 0.5em
  minus 0.4em\relax IEEE, 2016, pp. 1--6.

\bibitem{wymeersch20175g}
H.~Wymeersch, G.~Seco-Granados, G.~Destino, D.~Dardari, and F.~Tufvesson,
  ``{5G} {mmWave} positioning for vehicular networks,'' \emph{IEEE Wireless
  Communications}, vol.~24, no.~6, pp. 80--86, 2017.

\bibitem{choudhary2017distributed}
S.~Choudhary, L.~Carlone, C.~Nieto, J.~Rogers, H.~I. Christensen, and
  F.~Dellaert, ``Distributed mapping with privacy and communication
  constraints: Lightweight algorithms and object-based models,'' \emph{The
  International Journal of Robotics Research}, vol.~36, no.~12, pp. 1286--1311,
  2017.

\bibitem{lajoie2020door}
P.-Y. Lajoie, B.~Ramtoula, Y.~Chang, L.~Carlone, and G.~Beltrame,
  ``{DOOR-SLAM}: Distributed, online, and outlier resilient slam for robotic
  teams,'' \emph{IEEE Robotics and Automation Letters}, vol.~5, no.~2, pp.
  1656--1663, 2020.

\bibitem{groves2013height}
P.~D. Groves and Z.~Jiang, ``Height aiding, {$C/N_0$} weighting and consistency
  checking for {GNSS} {NLOS} and multipath mitigation in urban areas,''
  \emph{The Journal of Navigation}, vol.~66, no.~5, pp. 653--669, 2013.

\bibitem{wymeersch20185g}
H.~Wymeersch, N.~Garcia, H.~Kim, G.~Seco-Granados, S.~Kim, F.~Went, and
  M.~Fr{\"o}hle, ``{5G} {mmWave} downlink vehicular positioning,'' in
  \emph{2018 IEEE Global Communications Conference (GLOBECOM)}.\hskip 1em plus
  0.5em minus 0.4em\relax IEEE, 2018, pp. 206--212.

\bibitem{win2018foundations1}
M.~Z. Win, R.~M. Buehrer, G.~Chrisikos, A.~Conti, and H.~V. Poor, ``Foundations
  and trends in localization {Technologies—Part} {I} [scanning the issue],''
  \emph{Proceedings of the IEEE}, vol. 106, no.~6, pp. 1019--1021, 2018.

\bibitem{win2018foundations2}
M.~Z. Win, R.~M. Buehrer, G.~Chrisikos, A.~Conti, H.~V. Poor \emph{et~al.},
  ``Foundations and trends in localization technologies-part {II},'' 2018.

\bibitem{Fleury1999Channel}
B.~H. Fleury, M.~Tschudin, R.~Heddergott, D.~Dahlhaus, and K.~I. Pedersen,
  ``Channel parameter estimation in mobile radio environments using the {SAGE}
  algorithm,'' \emph{IEEE Journal on Selected Areas in Communications},
  vol.~17, no.~3, pp. 434--450, 1999.

\bibitem{richter2005estimation}
A.~Richter, ``Estimation of radio channel parameters: Models and
  algorithms.''\hskip 1em plus 0.5em minus 0.4em\relax ISLE, 2005.

\bibitem{3gpp_TS_36_211}
\BIBentryALTinterwordspacing
3GPP, ``{Physical channels and modulation (Release 15)},'' {3rd Generation
  Partnership Project (3GPP)}, Tech. Rep. 36.211, 2019, version 15.7.0.
  [Online]. Available:
  \url{https://www.3gpp.org/ftp/specs/archive/36_series/36.211/36211-f80.zip}
\BIBentrySTDinterwordspacing

\bibitem{3gpp_TS_38_211}
\BIBentryALTinterwordspacing
------, ``{NR; Physical channels and modulation},'' {3rd Generation Partnership
  Project (3GPP)}, Tech. Rep. 38.211, 2019, version 15.7.0. [Online].
  Available:
  \url{https://www.3gpp.org/ftp/specs/archive/38_series/38.211/38211-f70.zip}
\BIBentrySTDinterwordspacing

\bibitem{shahmansoori20155g}
A.~Shahmansoori, G.~E. Garcia, G.~Destino, G.~Seco-Granados, and H.~Wymeersch,
  ``{5G} position and orientation estimation through millimeter wave mimo,'' in
  \emph{2015 IEEE Globecom Workshops (GC Wkshps)}.\hskip 1em plus 0.5em minus
  0.4em\relax IEEE, 2015, pp. 1--6.

\bibitem{shahmansoori2018position}
------, ``Position and orientation estimation through millimeter-wave mimo in
  {5G} systems,'' \emph{IEEE Transactions on Wireless Communications}, vol.~17,
  no.~3, pp. 1822--1835, 2018.

\bibitem{del2017survey}
J.~A. del Peral-Rosado, R.~Raulefs, J.~A. L{\'o}pez-Salcedo, and
  G.~Seco-Granados, ``Survey of cellular mobile radio localization methods:
  From {1G} to {5G},'' \emph{IEEE Communications Surveys \& Tutorials},
  vol.~20, no.~2, pp. 1124--1148, 2017.

\bibitem{bulusu2000gps}
N.~Bulusu, J.~Heidemann, D.~Estrin \emph{et~al.}, ``{GPS-less} low-cost outdoor
  localization for very small devices,'' \emph{IEEE personal communications},
  vol.~7, no.~5, pp. 28--34, 2000.

\bibitem{win2018theoretical}
M.~Z. Win, Y.~Shen, and W.~Dai, ``A theoretical foundation of network
  localization and navigation,'' \emph{Proceedings of the IEEE}, vol. 106,
  no.~7, pp. 1136--1165, 2018.

\bibitem{Marano2010NLOS}
S.~Marano, W.~M. Gifford, H.~Wymeersch, and M.~Z. Win, ``{NLOS} identification
  and mitigation for localization,'' \emph{IEEE Journal on Selected Areas in
  Communications}, vol.~28, no.~7, pp. 1026--1035, 2010.

\bibitem{mingyang2008distributed}
S.~Mingyang, T.~Xiaofeng, X.~Yongtai, and H.~Xiao, ``A distributed
  multi-antenna based {NLOS} error elimination algorithm for mobile
  localization,'' in \emph{2008 4th IEEE International Conference on Circuits
  and Systems for Communications}.\hskip 1em plus 0.5em minus 0.4em\relax IEEE,
  2008, pp. 411--415.

\bibitem{dammann2015prospects}
A.~Dammann, R.~Raulefs, and S.~Zhang, ``On prospects of positioning in {5G},''
  in \emph{2015 IEEE International Conference on Communication Workshop
  (ICCW)}.\hskip 1em plus 0.5em minus 0.4em\relax IEEE, 2015, pp. 1207--1213.

\bibitem{koivisto2017joint}
M.~Koivisto, M.~Costa, J.~Werner, K.~Heiska, J.~Talvitie, K.~Lepp{\"a}nen,
  V.~Koivunen, and M.~Valkama, ``Joint device positioning and clock
  synchronization in {5G} ultra-dense networks,'' \emph{IEEE Transactions on
  Wireless Communications}, vol.~16, no.~5, pp. 2866--2881, 2017.

\bibitem{koivisto2017high}
M.~Koivisto, A.~Hakkarainen, M.~Costa, P.~Kela, K.~Leppanen, and M.~Valkama,
  ``High-efficiency device positioning and location-aware communications in
  dense {5G} networks,'' \emph{IEEE Communications Magazine}, vol.~55, no.~8,
  pp. 188--195, 2017.

\bibitem{savic2015fingerprinting}
V.~Savic and E.~G. Larsson, ``Fingerprinting-based positioning in distributed
  massive mimo systems,'' in \emph{2015 IEEE 82nd Vehicular Technology
  Conference (VTC2015-Fall)}.\hskip 1em plus 0.5em minus 0.4em\relax IEEE,
  2015, pp. 1--5.

\bibitem{leitinger2016belief}
E.~Leitinger, F.~Meyer, P.~Meissner, K.~Witrisal, and F.~Hlawatsch, ``Belief
  propagation based joint probabilistic data association for multipath-assisted
  indoor navigation and tracking,'' in \emph{2016 International Conference on
  Localization and GNSS (ICL-GNSS)}.\hskip 1em plus 0.5em minus 0.4em\relax
  IEEE, 2016, pp. 1--6.

\bibitem{leitinger2015evaluation}
E.~Leitinger, P.~Meissner, C.~R{\"u}disser, G.~Dumphart, and K.~Witrisal,
  ``Evaluation of position-related information in multipath components for
  indoor positioning,'' \emph{IEEE Journal on Selected Areas in
  communications}, vol.~33, no.~11, pp. 2313--2328, 2015.

\bibitem{gentner2016multipath}
C.~Gentner, T.~Jost, W.~Wang, S.~Zhang, A.~Dammann, and U.-C. Fiebig,
  ``Multipath assisted positioning with simultaneous localization and
  mapping,'' \emph{IEEE Transactions on Wireless Communications}, vol.~15,
  no.~9, pp. 6104--6117, 2016.

\bibitem{leitinger2015simultaneous}
E.~Leitinger, P.~Meissner, M.~Lafer, and K.~Witrisal, ``Simultaneous
  localization and mapping using multipath channel information,'' in \emph{2015
  IEEE International Conference on Communication Workshop (ICCW)}.\hskip 1em
  plus 0.5em minus 0.4em\relax IEEE, 2015, pp. 754--760.

\bibitem{leitinger2019belief}
E.~Leitinger, F.~Meyer, F.~Hlawatsch, K.~Witrisal, F.~Tufvesson, and M.~Z. Win,
  ``A belief propagation algorithm for multipath-based {SLAM},'' \emph{IEEE
  transactions on wireless communications}, 2019.

\bibitem{sorenson1971recursive}
H.~W. Sorenson and D.~L. Alspach, ``Recursive bayesian estimation using
  {Gaussian} sums,'' \emph{Automatica}, vol.~7, no.~4, pp. 465--479, 1971.

\bibitem{Casella1996Rao}
G.~Casella and C.~P. Robert, ``{Rao-Blackwellisation} of sampling schemes,''
  1996.

\bibitem{horiba2015accurate}
M.~Horiba, E.~Okamoto, T.~Shinohara, and K.~Matsumura, ``An accurate
  indoor-localization scheme with nlos detection and elimination exploiting
  stochastic characteristics,'' \emph{IEICE Transactions on Communications},
  vol.~98, no.~9, pp. 1758--1767, 2015.

\bibitem{jiao2009lcrt}
L.~Jiao, F.~Y. Li, and Z.~Xu, ``{LCRT}: A {TOA} based mobile terminal
  localization algorithm in {NLOS} environment,'' in \emph{VTC Spring 2009-IEEE
  69th Vehicular Technology Conference}.\hskip 1em plus 0.5em minus 0.4em\relax
  IEEE, 2009, pp. 1--5.

\bibitem{witrisal2016high}
K.~Witrisal, P.~Meissner, E.~Leitinger, Y.~Shen, C.~Gustafson, F.~Tufvesson,
  K.~Haneda, D.~Dardari, A.~F. Molisch, A.~Conti \emph{et~al.}, ``High-accuracy
  localization for assisted living: {5G} systems will turn multipath channels
  from foe to friend,'' \emph{IEEE Signal Processing Magazine}, vol.~33, no.~2,
  pp. 59--70, 2016.

\bibitem{wang2011omnidirectional}
Z.~Wang and S.~A. Zekavat, ``Omnidirectional mobile nlos identification and
  localization via multiple cooperative nodes,'' \emph{IEEE Transactions on
  Mobile Computing}, vol.~11, no.~12, pp. 2047--2059, 2011.

\bibitem{han2018hidden}
K.~Han, S.-W. Ko, H.~Chae, B.-H. Kim, and K.~Huang, ``Hidden vehicles
  positioning via asynchronous {V2V} transmission: A multi-path-geometry
  approach,'' \emph{arXiv preprint arXiv:1804.10778}, 2018.

\bibitem{mendrzik2018joint}
R.~Mendrzik, H.~Wymeersch, and G.~Bauch, ``Joint localization and mapping
  through millimeter wave {MIMO} in {5G} systems,'' in \emph{2018 IEEE Global
  Communications Conference (GLOBECOM)}.\hskip 1em plus 0.5em minus 0.4em\relax
  IEEE, 2018, pp. 1--6.

\bibitem{palacios2017jade}
J.~Palacios, P.~Casari, and J.~Widmer, ``{JADE}: Zero-knowledge device
  localization and environment mapping for millimeter wave systems,'' in
  \emph{IEEE INFOCOM 2017-IEEE Conference on Computer Communications}.\hskip
  1em plus 0.5em minus 0.4em\relax IEEE, 2017, pp. 1--9.

\bibitem{palacios2018communication}
J.~Palacios, G.~Bielsa, P.~Casari, and J.~Widmer, ``Communication-driven
  localization and mapping for millimeter wave networks,'' in \emph{IEEE
  INFOCOM 2018-IEEE Conference on Computer Communications}.\hskip 1em plus
  0.5em minus 0.4em\relax IEEE, 2018, pp. 2402--2410.

\bibitem{aladsani2019leveraging}
M.~Aladsani, A.~Alkhateeb, and G.~C. Trichopoulos, ``Leveraging {mmWave}
  imaging and communications for simultaneous localization and mapping,'' in
  \emph{ICASSP 2019-2019 IEEE International Conference on Acoustics, Speech and
  Signal Processing (ICASSP)}.\hskip 1em plus 0.5em minus 0.4em\relax IEEE,
  2019, pp. 4539--4543.

\bibitem{yassin2018mosaic}
A.~Yassin, Y.~Nasser, A.~Y. Al-Dubai, and M.~Awad, ``Mosaic: Simultaneous
  localization and environment mapping using {mmWave} without a-priori
  knowledge,'' \emph{IEEE Access}, vol.~6, pp. 68\,932--68\,947, 2018.

\bibitem{kim20205g}
H.~Kim, K.~Granstr{\"o}m, L.~Gao, G.~Battistelli, S.~Kim, and H.~Wymeersch,
  ``{5G} {mmWave} cooperative positioning and mapping using multi-model phd
  filter and map fusion,'' \emph{IEEE Transactions on Wireless Communications},
  2020.

\bibitem{blanco2019performance}
A.~Blanco, N.~Ludant, S.~Zhenyu, W.~Yi, and J.~Widmer, ``Performance evaluation
  of single base station {ToA-AoA} localization in an {LTE} testbed,'' 2019.

\bibitem{bresson2017simultaneous}
G.~Bresson, Z.~Alsayed, L.~Yu, and S.~Glaser, ``Simultaneous localization and
  mapping: A survey of current trends in autonomous driving,'' \emph{IEEE
  Transactions on Intelligent Vehicles}, vol.~2, no.~3, pp. 194--220, 2017.

\bibitem{durrant2006simultaneous}
H.~Durrant-Whyte and T.~Bailey, ``Simultaneous localization and mapping: part
  {I},'' \emph{IEEE robotics \&amp; automation magazine}, vol.~13, no.~2, pp.
  99--110, 2006.

\bibitem{dissanayake2001solution}
M.~G. Dissanayake, P.~Newman, S.~Clark, H.~F. Durrant-Whyte, and M.~Csorba, ``A
  solution to the simultaneous localization and map building {(SLAM)}
  problem,'' \emph{IEEE Transactions on robotics and automation}, vol.~17,
  no.~3, pp. 229--241, 2001.

\bibitem{mullane2011random}
J.~Mullane, B.-N. Vo, M.~D. Adams, and B.-T. Vo, ``A random-finite-set approach
  to bayesian {SLAM},'' \emph{IEEE Transactions on Robotics}, vol.~27, no.~2,
  pp. 268--282, 2011.

\bibitem{frey2007clustering}
B.~J. Frey and D.~Dueck, ``Clustering by passing messages between data
  points,'' \emph{science}, vol. 315, no. 5814, pp. 972--976, 2007.

\bibitem{liu2017distributed}
W.~Liu, G.~Qin, Y.~He, and F.~Jiang, ``Distributed cooperative reinforcement
  learning-based traffic signal control that integrates {V2X} networks’
  dynamic clustering,'' \emph{IEEE Transactions on Vehicular Technology},
  vol.~66, no.~10, pp. 8667--8681, 2017.

\bibitem{han2019production}
Y.~Han, H.~Wu, M.~Jia, Z.~Geng, and Y.~Zhong, ``Production capacity analysis
  and energy optimization of complex petrochemical industries using novel
  extreme learning machine integrating affinity propagation,'' \emph{Energy
  Conversion and Management}, vol. 180, pp. 240--249, 2019.

\bibitem{van2001unscented}
R.~Van Der~Merwe, A.~Doucet, N.~De~Freitas, and E.~A. Wan, ``The unscented
  particle filter,'' in \emph{Advances in neural information processing
  systems}, 2001, pp. 584--590.

\bibitem{Siciliano2016Robotics}
B.~Siciliano and O.~Khatib, \emph{Robotics and the Handbook}.\hskip 1em plus
  0.5em minus 0.4em\relax Springer, 2016.

\bibitem{gustafsson2002particle}
F.~Gustafsson, F.~Gunnarsson, N.~Bergman, U.~Forssell, J.~Jansson, R.~Karlsson,
  and P.-J. Nordlund, ``Particle filters for positioning, navigation, and
  tracking,'' \emph{IEEE Transactions on signal processing}, vol.~50, no.~2,
  pp. 425--437, 2002.

\bibitem{shapiro2003monte}
A.~Shapiro, ``{Monte} {Carlo} sampling methods,'' \emph{Handbooks in operations
  research and management science}, vol.~10, pp. 353--425, 2003.

\bibitem{yin2018gnss}
L.~Yin, Q.~Ni, and Z.~Deng, ``A {GNSS}/{5G} integrated positioning methodology
  in {D2D} communication networks,'' \emph{IEEE Journal on Selected Areas in
  Communications}, vol.~36, no.~2, pp. 351--362, 2018.

\bibitem{mendrzik2019enabling}
R.~Mendrzik, F.~Meyer, G.~Bauch, and M.~Z. Win, ``Enabling situational
  awareness in millimeter wave massive mimo systems,'' \emph{IEEE Journal of
  Selected Topics in Signal Processing}, vol.~13, no.~5, pp. 1196--1211, 2019.

\end{thebibliography}

\end{document}